\documentclass{achemso}

\usepackage[version=3]{mhchem} 
\usepackage{hyperref}       
\usepackage{url}            
\usepackage{booktabs}       
\usepackage{amsfonts}       
\usepackage{nicefrac}       
\usepackage{microtype}      
\usepackage{bm}             
\usepackage{multirow}       
\usepackage{indentfirst}
\usepackage{threeparttable}
\graphicspath{{media/}}     

\author{Yicheng Chen}
\altaffiliation{These authors contribute equally.}
\author{Wenjie Yan}
\altaffiliation{These authors contribute equally.}
\author{Zhanfeng Wang}
\author{Jianming Wu}
\email{jianmingwu@fudan.edu.cn}
\author{Xin Xu}
\email{xxchem@fudan.edu.cn}
\affiliation[Fudan University]
{Department of Chemistry, Collaborative Innovation Center of Chemistry for Energy Materials, Shanghai Key Laboratory of Molecular Catalysis and Innovative Materials, MOE Key Laboratory of Computational Physical Sciences, Fudan University, 200433, Shanghai, People's Republic of China}
\alsoaffiliation[Hefei National Labtoratory]
{Hefei National Laboratory, 230088, Hefei, People's Republic of China}

\title{Constructing accurate and efficient general-purpose atomistic machine learning model with transferable accuracy for quantum chemistry}


\begin{document}

\begin{abstract}
Density Functional Theory (DFT) has been a cornerstone in computational science, providing powerful insights into structure-property relationships for molecules and materials through first-principles quantum-mechanical (QM) calculations. However, the advent of atomistic machine learning (ML) is reshaping the landscape by enabling large-scale dynamics simulations and high-throughput screening at DFT-equivalent accuracy with drastically reduced computational cost. Yet, the development of general-purpose atomistic ML models as surrogates for QM calculations faces several challenges, particularly in terms of model capacity, data efficiency, and transferability across chemically diverse systems. This work introduces a novel extension of the polarizable atom interaction neural network (namely, XPaiNN) to address these challenges. Two distinct training strategies have been employed, one direct-learning and the other $\Delta$-ML on top of a semi-empirical QM method. These methodologies have been implemented within the same framework, allowing for a detailed comparison of their results. The XPaiNN models, in particular the one using $\Delta$-ML, not only demonstrate competitive performance on standard benchmarks, but also demonstrate the effectiveness against other ML models and QM methods on comprehensive downstream tasks, including non-covalent interactions, reaction energetics, barrier heights, geometry optimization and reaction thermodynamics, etc. This work represents a significant step forward in the pursuit of accurate and efficient atomistic ML models of general-purpose, capable of handling complex chemical systems with transferable accuracy.
\end{abstract}

\section{Introduction}
Density Functional Theory (DFT)\cite{hohenberg1964dft,kohn1965kohnsham}, stands out as an outstanding representative of modern electronic structure methods, being indispensable across various scientific disciplines over the past decades, owing to its predictive power in elucidating and predicting structure-property relationships of molecules and materials through quantum-mechanical (QM) calculations at the atomic level\cite{marzari2021electronic}. However, the emergence of atomistic machine learning (ML) approaches\cite{butler2018mlreview, margraf2023scidriven} is transforming the research paradigm. These atomistic ML models directly map geometric inputs of molecules or materials to the respective target properties, enabling large-scale dynamics simulations\cite{unke2021MLFF} and high-throughput screening\cite{merchant2023GNoME} with accuracy comparable to those of DFT calculations, all while bypassing the traditionally intensive and costly routines of DFT calculations. Once these models are trained on datasets encompassing the relevant chemical space for the systems of interest, they significantly enhance efficiency and predictive power.

Despite the successes that have already been achieved, developing generally applicable atomistic ML models as surrogates for QM calculations remains a significant challenge\cite{poltavsky_mlff-perspective_2021}. Specifically, model capacity and data efficiency are the major concerns. A lower capacity model is unable to learn and capture complex relationships or patterns within the data, whereas a higher capacity model may perform exceptionally well on the training data but poorly on unseen or test data. As accessing vast quantities of high-quality data for achieving comprehensive coverage of all chemical space is inherently constrained, data efficiency is crucial for reducing the dependence on large annotated datasets and for developing ML models with high transferable accuracy. Furthermore, in practical applications, chemists are more interested in relative energy changes that occur during dynamic processes or chemical reactions, as opposed to mere deviations observed in static systems. This highlights a clear disparity between methodological advancements and practical needs, where the models' ability to generalize from their training samples to chemically distinct systems, especially in downstream tasks involving relative energetics, is notably problematic\cite{margraf2023scidriven}.

Both Deep Neural Networks (DNNs)\cite{ani, ani-1ccx, ani-2x, aiqm1} and Graph Neural Networks (GNNs)\cite{qiao2020orbnet, christensen2021orbnetdenali, orbnet-equi} have been utilized in constructing atomistic ML models, aiming as general alternatives for routine computational tasks in quantum chemistry. Currently, AIQM1\cite{aiqm1} and OrbNet-Equi\cite{orbnet-equi} represent the state-of-the-art (SOTA) for two types of networks, without or with, respectively, integrating electronic structure information within the NN framework. Despite their remarkable performances, there remains room for improvement in their general applicability at this stage, primarily due to architectural limitations of the models. For instance, AIQM1 is limited to systems made solely from C, H, N, and O atoms; whereas obtaining higher-order analytical gradients with OrbNet-Equi is not straightforward\cite{qiao2020orbnetgradients}.

In this work, we propose a novel design of equivariant GNN, aiming to construct an accurate and efficient atomistic ML framework for broad applicability. We dub this model XPaiNN, which architecturally extends the successful \textit{Polarizable Atom Interaction Neural Network} (PaiNN)\cite{painn}. Comprehensive details regarding the model framework are presented in the \textbf{Methods} section and the \textbf{Supporting Information (SI)}. XPaiNN has demonstrated competitive performance alongside the existing ML models on standard benchmark dataset of QM9\cite{qm9}, indicative of its model capacity to cover a wide chemical space. Furthermore, we have trained XPaiNN using the SPICE dataset\cite{spice}, containing one million conformers across ten different element types, to establish NN potential energy models that are generally useful for organic molecules. Two distinct training methodologies have been employed, one direct-learning and the other $\Delta$-ML on top of a semi-empirical QM (SQM) method, yielding two models that have been compared in detail against other ML models and QM methods in downstream applications. These tasks include energy benchmarks for reactions, assessments of non-covalent interactions (NCIs), conformational analyses, barrier height determinations, as well as geometry optimizations and reaction thermodynamics, essentially covering all kinds of routine calculations for thermochemistry. This is the first time these two training methodologies have been directly compared within the same framework, providing valuable insights into their relative performance and applicability. The outcomes have robustly validated our strategy and underscored the practical utility of our models. Significantly, our results have demonstrated that the integration of XPaiNN with the SQM method GFN2-xTB\cite{gfn2-xtb} yields a highly promising general-purpose atomistic ML model, namely XPaiNN@xTB, that boosts not only high accuracy but also transferable performance across diverse chemical contexts.

\section{Construction of XPaiNN models}
\subsection{Related works}
By treating atomic systems as graph-structured data, GNNs have demonstrated remarkable performance in end-to-end atomistic ML applications\cite{reiser2022graph}. The inclusion of rotation and inversion equivariance has led to the advent of equivariant GNNs\cite{thomas2018TFN, painn, nequip, e3nn_paper, equiformer, orbnet-equi}, which have made significant strides in virtually every aspect when compared to their invariant equivalents\cite{schnet, dimenet++}. Notably, equivariant GNN models have been reported to exhibit superior data efficiency in achieving comparable (or even better) levels of accuracy\cite{nequip}. An overview of group equivariance and equivariant GNNs is provided in the \textbf{SI}.

One primary critique of equivariant GNNs lies in the computationally intensive tensor product operations for generating equivariant hidden features, which notably impairs model efficiency\cite{equiformer}. Acknowledging this limitation, Sch\"{u}tt and co-workers introduced the PaiNN model\cite{painn}, which circumvents this problem of tensor contraction convolutions. PaiNN preserves equivariance in the neural network by implementing separate transformation rules for scalar and Cartesian vector feature channels. TensorNet\cite{tensornet}, another equivariant GNN, adheres to a similar design principle, focusing on the generation and transformation of different components of rank-2 tensors in Cartesian representation, thereby achieving improved performance over PaiNN.

The approach of $\Delta$-ML\cite{ramakrishnan2015deltaML, wu2007x1, zhang2010x1, zhou2016X3D}, has long been pivotal in enhancing the transferability of atomistic ML models, where a cheaper method, often DFT, is employed as the baseline method for calibrating the target property with higher accuracy that would otherwise be attainable only with Coupled-Cluster Singles Doubles with perturbative Triples (CCSD(T)), or with high-quality experiment results. The renaissance of SQM methods represents a recent direction\cite{christensen2016SQM, dral2024SQM}, which enables computational speeds several orders of magnitude faster compared to conventional DFT calculations, albeit with a trade-off involving reduced accuracy. As systematic approximations are incorporated, these SQMs are drawing new attention as the baseline method in $\Delta$-ML models\cite{wengert2022hybrid, fan2022SQM}.

There are two prominent series of atomistic ML methods, aiming for transferable and accurate modeling of general-purpose. One belongs to the ANI potential family\cite{ani, ani-1ccx, ani-2x, aiqm1}, developed using the classical framework of Behler-Parrinello neural network\cite{behler2007BPNN}. The other key player is the OrbNet family\cite{qiao2020orbnet, christensen2021orbnetdenali, orbnet-equi}, introduced by Miller's group, which embodies a category of QM-informed ML models, using features, such as the single-particle Hamiltonian\cite{welborn2018QMinformed}, generated from QM calculations as inputs for GNNs\cite{welborn2018QMinformed, cheng2019QMinformed, cheng2022molecular}. Of note, AIQM1\cite{aiqm1} and OrbNet-Equi\cite{orbnet-equi} represent the SOTA with remarkable accuracy in chemical benchmark tests. Both of these models adopt the $\Delta$-ML strategy, wherein SQM methods serve as the baseline methods.

\subsection{XPaiNN architecture}
\begin{figure}[htbp]
    \centering
    \includegraphics[width=0.7\textwidth]{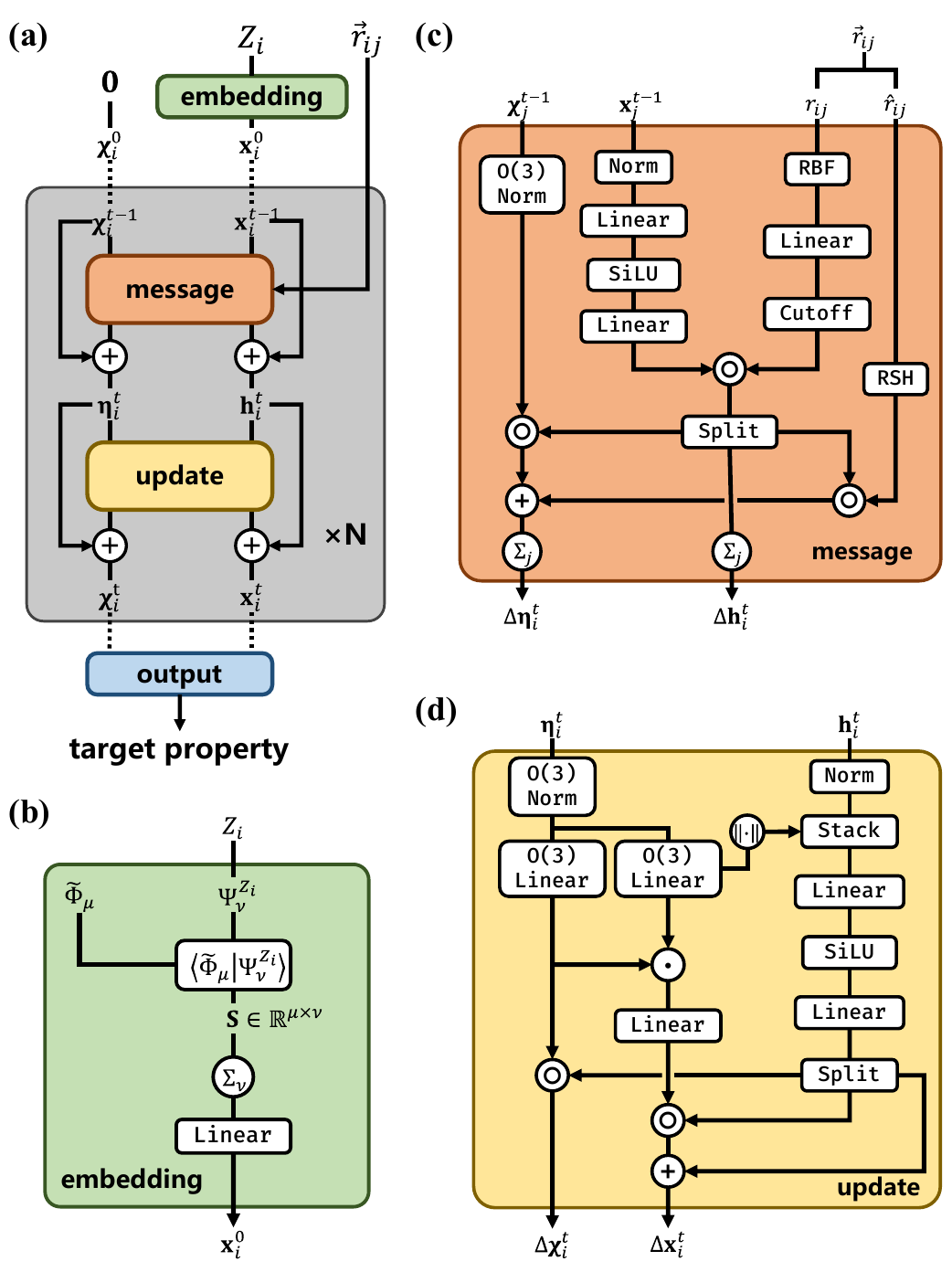}
    \caption{
        \textbf{Schematic diagram of XPaiNN model architecture.}        
        \textbf{(a)} Architecture overview.
        \textbf{(b)} Atom embedding layer.
        \textbf{(c)} Massage passing layer.
        \textbf{(d)} Node update layer.
        }
    \label{fig:struct}
\end{figure}

The architecture of XPaiNN is illustrated in Figure \ref{fig:struct}. As shown, XPaiNN inherits its basic structure from PaiNN, consisting of three primary components: \textbf{Embedding}, \textbf{Message Passing Convolution} (involving both \textbf{Message Passing} and \textbf{Node Updating}), and the \textbf{Output} block. In order to bolster the model capacity, one major change to the original PaiNN framework is to substitute the vector feature channel $\mathbf{v}$ in PaiNN with a spherical feature channel $\boldsymbol{\chi}$ in XPaiNN, an innovation explained further in subsequent sections. Moreover, we have devised a novel set of element-wise scalar embeddings for initializing node features within graph data, intended to reflect the periodic table trend for properties of chemical elements. Detailed descriptions regarding the remaining of the model architecture are provided in the \textbf{SI}.

\textbf{Embedding}.We employ a set of precomputed element-specific features $\mathbf{e}^{Z_A}$ for each atomic species A, identified by its atomic number $Z_A$, which differs from the conventional one-hot embedding typically used in atomistic ML models. These embedding features are linearly transformed to the hidden node features $\mathbf{x}_i^0$, serving as the initial input to the scalar feature channel of XPaiNN for the $i$-th node. Additionally, the input to the spherical feature channel, $\boldsymbol{\chi}_i^0$ is initialized with all zeros.

\begin{equation}
    \mathbf{x}_i^0 = \mathbf{e}^{Z_i} \mathbf{W}^\top + \mathbf{b}
\end{equation}
\begin{equation}
    \boldsymbol{\chi}_i^0 = \bigoplus_{l} \boldsymbol{\omega}^{(l)}_{i} =\boldsymbol{0}
    \label{equ:sphten}
\end{equation}

\textbf{Message Passing}. The improvement of TensorNet, as compared to PaiNN, illustrates the significance of higher $l$-order feature channels in enhancing the model capacity of an equivariant GNN method. Meanwhile it is much easier to handle $l$-order channels in spherical representation than in Cartesian form. Consequently, we extend the original vector feature channel $\mathbf{v}_i$ ($l=1$) of PaiNN to a spherical feature channel $\boldsymbol{\chi}_i$ of XPaiNN. As shown in Eq.(\ref{equ:sphten}), $\boldsymbol{\chi}_i$ is constituted by the direct sum of irreducible representations (Irreps) $\boldsymbol{\omega}^{(l)}_{i}$ corresponding to angular momentum number $l$, which is a hyperparameter of great significance in XPaiNN, offering flexibility in dimension adjustment. This strategy amplifies model capacity desirably, while only moderately increases the number of trainable parameters, thereby preserving the computational efficiency inherent to PaiNN. To incorporate this change in the feature channels, we use an equivariant filter by expanding the inter-atomic directions $\hat{\mathbf{r}}_{ij}$ with real spherical harmonics (RSH) in the \textbf{Message Passing} layer. This filter is responsible for directly generating equivariant features of the required order and shape. The subsequent message passing routine mirrors that of PaiNN, where both scalar and spherical features are updated through aggregation over transformed features of neighboring atoms $\mathcal{N}(i)$, which can be expressed as
\begin{equation}
    \mathbf{h}_i^t = \mathbf{x}_i^{t-1} + 
        \sum_{j \in \mathcal{N} \left( i \right)}
            f_{m}^t\left( \mathbf{x}_j^{t-1}, \mathbf{r}_{ij} \right)
\end{equation}
\begin{equation}
    \boldsymbol{\eta}_i^t = \boldsymbol{\chi}_i^{t-1} +
        \sum_{j \in \mathcal{N} \left( i \right)}
            \phi_{m}^t\left( \mathbf{x}_j^{t-1}, \boldsymbol{\chi}_j^{t-1}, \mathbf{r}_{ij} \right)
\end{equation}
where the concrete operations of message functions $f_m$ and $\phi_m$ are illustrated in Figure \ref{fig:struct}\textbf{(c)}.

\subsection{Datasets and training}
The QM9 dataset is widely used for assessing the performance of atomistic ML models in predicting molecular properties. It contains 133,885 small organic molecules at the respective DFT-optimized geometry, with target properties computed at the same level\cite{qm9}. To benchmark the model capacity of the XPaiNN framework, we also employ the QM9 dataset\cite{qm9}, capitalizing on existing comparisons among various atomistic ML models.

Adequate training data is a prerequisite for the development of a general-purpose potential energy surface model. Notably, most of the ANI potentials are developed based on the ANI-1x dataset\cite{smith2020ani-1-dataset}, encompassing 5 million single organic molecules composed of C, H, N, and O. Conversely, the OrbNet Denali dataset\cite{christensen2021orbnetdenali} contains 2.3 million molecular conformations covering 17 elements, with a subset serving as the training set for OrbNet-Equi. Furthermore, we believe that central to constructing an accurate atomistic ML model of practical value are considerations regarding the breadth of chemical space coverage and the reliability of DFT-calculated labels.

With these considerations in mind, this work utilizes the SPICE dataset\cite{spice} as its foundation. Comprising up to 1.1 million configurations of small organic molecules, dimers, dipeptides, and solvated amino acids, this dataset covers a wide range of covalent interactions, as well as NCIs in main-group chemistry. It furnishes total energies, forces, and additional properties computed at the $\omega$B97M-D3(BJ)\cite{wb97m-v,wb97m-d3}/def2-TZVPPD level of theory, which, to the best of our knowledge, is the first QM dataset of this size with such an accuracy. SPICE further ensures comprehensive coverage of relevant conformational landscapes for each sampled molecule, with guidelines conducive to further data augmentation\cite{spice}. Thus, this dataset is anticipated to be an ideal choice for training accurate potential energy model for general-purpose in a data-efficient way. The training set used in this work is a curated subset of SPICE, including only neutral, closed-shell molecular systems featuring H, C, N, O, F, P, S, Cl, Br, and I (see the \textbf{SI} for more details).

On top of this selected SPICE dataset, we trained two models: one that directly fits target labels of energies and forces, and the other that adopts the $\Delta$-ML strategy, using the SQM method GFN2-xTB\cite{gfn2-xtb} as the baseline method. Consequently, the energy and force evaluations by the latter model, XPaiNN@xTB, can be decomposed into two parts:
\begin{equation}
    E_{total}^{\mathrm{XPaiNN@xTB}} = E^{\mathrm{XPaiNN}} + E^{\mathrm{GFN2\text{-}xTB}}
\end{equation}
\begin{equation}
    \vec{F}^{\mathrm{XPaiNN@xTB}}_{A} = \vec{F}^{\mathrm{XPaiNN}}_{A} + \vec{F}^{\mathrm{GFN2\text{-}xTB}}_{A}
\end{equation}

Training for this model focuses on the residuals between outputs of the baseline method and those of the target level of theory, with other hyper-parameters and training strategies remaining consistent as the direct-learning model. This setup enables a clear illustration of the advantages and drawbacks between direct-learning and the $\Delta$-ML strategy through comparative testing performance. Detailed training methodologies for these models are outlined in the \textbf{SI}, along with hyper-parameters provided in Tables S1-S3.

\section{Results}
\subsection{Performance on QM9}
The results of test performance for XPaiNN across the 12 prediction tasks within the QM9 dataset are summarized in Table \ref{tab:qm9}, along with those reported by other GNN and equivariant GNN models. As evident, XPaiNN demonstrates low test errors across all tasks, positioning it amongst the top performers against the QM9 benchmarking. Notably, substantial and consistent advancements are discernible when contrasting the results of XPaiNN with those of PaiNN. Specifically, in properties pertaining to frontier orbitals, namely, $\epsilon_{\text{HOMO}}$, $\epsilon_{\text{LUMO}}$ and $\Delta\epsilon$, the mean absolute errors (MAEs) in these subsets are reduced by an average of 18\%. With respect to thermodynamic properties, including $U_0$, $U$, $H$ and $G$, an additional improvement averaging 0.4 meV is accomplished, a significant refinement given the already insignificant error margins. Further ablation studies, as detailed in Table S4 in the \textbf{SI}, suggest that these improvements predominantly derive from the incorporation of higher-order features in the spherical channels of XPaiNN, demonstrating the success of our model architecture.

\begin{table}[htbp]
    \centering
    \resizebox{\textwidth}{!}{
        \begin{tabular}{llllllll}
            \toprule
            Task & Unit & SchNet\textsuperscript{\emph{b}} & DimeNet++\textsuperscript{\emph{b}} & PaiNN\textsuperscript{\emph{b}} & SphereNet\textsuperscript{\emph{b}} & Equiformer\textsuperscript{\emph{b}} & XPaiNN\textsuperscript{\emph{c}} \\
            \midrule
            $\mu$ & D & 0.033 & 0.030 & 0.012 & 0.024 & \textbf{0.011} & \textbf{0.011±0.000} \\  
            $\alpha$ & $a_0^3$ & 0.235 & 0.044 & 0.045 & 0.045 & \textbf{0.023} & 0.042±0.000 \\
            $\epsilon_{\text{HOMO}}$ & meV & 41 & 24.6 & 27.6 & 22.8 & \textbf{15} & 21.7±0.2\\
            $\epsilon_{\text{LUMO}}$ & meV & 34 & 19.5 & 20.4 & 18.9 & \textbf{14} & 16.4±0.2 \\
            $\Delta\epsilon$ & meV & 63 & 32.6 & 45.7 & 31.1 & \textbf{30} & 38.1±0.7\\
            $\left<R^2\right>$ & $a_0^2$ & 0.073 & 0.331 & 0.066 & 0.268 & 0.251 & \textbf{0.065±0.002} \\
            zpve & meV & 1.7 & 1.21 & 1.28 & \textbf{1.12} & 1.26 & 1.20±0.01\\
            $U_0$ & meV & 14 & 6.32 & 5.85 & 6.26 & 6.59 & \textbf{5.43±0.08}\\
            $U$ & meV & 19 & 6.28 & 5.83 & 6.36 & 6.74 & \textbf{5.53±0.03}\\
            $H$ & meV & 14 & 6.53 & 5.98 & 6.33 & 6.63 & \textbf{5.41±0.03} \\
            $G$ & meV & 14 & 7.56 & 7.35 & 7.78 & 7.63 & \textbf{6.85±0.06}\\
            $C_v$ & cal/mol/K & 0.033 & 0.023 & 0.024 & \textbf{0.0215} & 0.023 & 0.023±0.000 \\
            \bottomrule
        \end{tabular}
    }
    \caption{Mean absolute errors on QM9 test set.\textsuperscript{\emph{a}}} 
    \begin{tablenotes}
        \item[1] \textsuperscript{\emph{a}} Best results in \textbf{bold};
        \item[2] \textsuperscript{\emph{b}} Results are taken from Ref.~\citenum{equiformer};
        \item[3] \textsuperscript{\emph{c}} Averaged over four models.
    \end{tablenotes}
    \label{tab:qm9}
\end{table}

Results summarized in Table S5 of the \textbf{SI} uncover a further dramatic improvement over XPaiNN on thermodynamic properties with XPaiNN@xTB, demonstrating the enhanced model capacity and data efficiency of the $\Delta$-ML strategy.

\subsection{Performance on downstream tasks}
To explore model transferability, we apply the two potential energy models, XPaiNN and XPaiNN@xTB, to a variety of downstream tasks, involving the calculations of relative energetic properties such as reaction energies, barrier heights, and NCIs, as well as geometry optimizations and reaction thermodynamics calculations.

For comparison, we consider three SOTA atomistic ML models, ANI-1ccx\cite{ani-1ccx}, AIQM1\cite{aiqm1}, and OrbNet-Equi\cite{orbnet-equi}, where their performances on these benchmarks are documented. Furthermore, the results from some recent SQM methods, i.e., GFN-xTB\cite{gfn-xtb}, GFN2-xTB\cite{gfn2-xtb}, and ODM2\cite{odm2}, are also included, as they served as the baseline methods for OrbNet-Equi, XPaiNN@xTB, and AIQM1, respectively. Additionally, we present results from the DFT method $\omega$B97M-D3(BJ)\cite{wb97m-v, wb97m-d3}, which represents the target level of theory for our models. It is important to acknowledge that the accuracy and transferability of these ML models vary as a result of differences in model architecture and training regimens, reflecting model capacity and data efficiency. Specifically, the ANI models, ANI-1ccx and AIQM1, have been trained on energies and forces at the $\omega$B97X\cite{wb97x} level and are subsequently fine-tuned via transfer learning to approximate the CCSD(T)/CBS level for energy predictions, albeit being limited to compounds comprising H, C, N, and O\cite{ani-1ccx, aiqm1}. Meanwhile, OrbNet-Equi has been trained on energy labels at the $\omega$B97X-D3/def2-TZVP level, covering a broader range of chemical elements\cite{orbnet-equi}.

Furthermore, we apply the two models, XPaiNN and XPaiNN@xTB, to reaction enthalpy calculations, wherein second-order derivatives of energy with respect to atomic positions are necessitated by the computations of vibrational frequencies and then thermo-corrections. This validates the potential of our models as a viable alternative for routine QM calculations of reaction thermodynamics, aiming to offer both high efficiency and accuracy.

\textbf{GMTKN55 benchmark}. To assess the models' performance in predicting relative energies, we examine XPaiNN and XPaiNN@xTB utilizing GMTKN55\cite{gmtkn55}, which is a benchmark widely acknowledged for its effectiveness in evaluating electronic structure methods. This dataset contains 55 subsets designed to evaluate various aspects of comprehensive main-group chemistry, targeting General Main-group Thermochemistry, Kinetics, and Noncovalent interactions\cite{gmtkn55}. It employs reference values often calculated at the gold-standard CCSD(T)/CBS level of theory. Our focus lies on subsets composed exclusively of neutral, closed-shell molecules, comprising elements of H, C, N, O, F, P, S, Cl, Br, and I, in accordance with the design of XPaiNN. Consequently, from the subsets within GMTKN55, 21 have been selectively considered, and the overall error analyses are summarized in Table \ref{tab:gmtkn55}. In order to quantify subset-specific errors, we employ the Weighted Total Mean Absolute Deviation of the second type (WTMAD-2), as defined in Ref.~\citenum{orbnet-equi}. The detailed formula is outlined in the \textbf{Computational Details} section in \textbf{Material and Methods}. Illustrations of the chosen subsets are provided in Table S6 in the \textbf{SI}, with further specifics available elsewhere\cite{gmtkn55}.

\begin{table}[htbp]
    \renewcommand{\arraystretch}{1.3}
    \centering
    \resizebox{\textwidth}{!}{
        \begin{tabular}{lcccccccc}
        \toprule
         & $\omega$B97M-D3(BJ) & GFN-xTB & GFN2-xTB & AIQM1 & OrbNet-Equi & XPaiNN & XPaiNN@xTB \\
        \midrule
        \multicolumn{8}{c}{Basic properties and reaction energies for small systems} \\
        FH51 & 1.86 & 22.09 & 20.92 & - & 9.40 & 8.25 & \textbf{6.80} \\
        TAUT15 & 5.41 & 108.14 & 18.31 & - & 21.78  &\textbf{7.81} & 8.18 \\
        \multicolumn{8}{c}{Reaction energies for large systems and isomerization reactions} \\
        DARC & 1.88 & 27.70 & 31.10 & \textbf{1.11} & 1.32 &2.73 & 1.85 \\
        BSR36 & 6.55 & 8.22 & 9.69 & \textbf{3.71} & 24.07 & 4.13 & 5.94 \\
        CDIE20 & 8.91 & 28.63 & 25.26 & - & 16.00 & 18.85 & \textbf{11.74} \\
        ISO34 & 2.38 & 24.56 & 26.93 & \textbf{1.70} & 7.28  &3.30 & 2.07 \\
        C60ISO & 7.90 & 4.56 & 3.36 & 27.71 & \textbf{2.32} & 23.98 & 5.09 \\
        \multicolumn{8}{c}{Reaction barrier heights} \\
        BHROT27 & 2.12 & 21.55 & 10.59 & - & 5.40  &6.72 & \textbf{3.57} \\
        PX13 & 2.69 & 14.14 & \textbf{4.66}& - & 22.08  &50.72 & 20.26\\
        WCPT18 & 1.81 & 8.61 & 6.24 & - & 12.10 & 10.08 & \textbf{3.97} \\
        \multicolumn{8}{c}{Intermolecular noncovalent interactions} \\
        ADIM6 & 6.82 & 17.06 & 19.48 & 14.97 & \textbf{4.64} & 4.93 & 8.55 \\
        S22 & 1.97 & 10.36 & 5.90 & 6.76 & 4.05 & 4.71 & \textbf{2.07} \\
        S66 & 2.15 & 11.16 & 7.60 & 6.45 & 5.16 & 4.12 & \textbf{2.91} \\
        HAL59 & 5.51 & 16.64 & \textbf{15.79} & - & 33.50 & 90.46 & 57.08 \\
        \multicolumn{8}{c}{Intramolecular noncovalent interactions} \\
        IDISP & 5.70 & 26.08 & 27.09 & 13.24 & 19.66 & 9.50 & \textbf{7.12} \\
        ACONF & 3.92 & 20.52 & 5.97 & 8.07 & \textbf{1.64} & 5.55 & 2.66 \\
        Amino20x4 & 6.12 & 25.97 & 22.23 & - & 7.01 & 13.85 & \textbf{6.99} \\
        PCONF21 & 11.76 & 76.04 & 61.62 & 27.38 & \textbf{18.00} & 33.25 & 20.25 \\
        MCONF & 2.46 & 16.51 & 19.70 & 4.13 & 5.39 & 9.90 & \textbf{4.31} \\
        SCONF & 1.81 & 30.93 & 20.30 & 11.68 & \textbf{6.59} & 13.86 & 9.18 \\
        BUT14DIOL & 4.11 & 19.36 & 25.36 & 9.17 & 12.02 & 3.78 & \textbf{2.56} \\
        \midrule
        WTMAD-2 (HCNO)\textsuperscript{\emph{b}} & 3.81 & 19.60 & 19.71 & 7.84 & 9.35 & 7.46 & \textbf{4.64} \\
        WTMAD-2 (Total)\textsuperscript{\emph{c}} & 4.15 & 22.41 & 18.45 & - & 11.77 & 17.20 & \textbf{10.44} \\
        \bottomrule
        \end{tabular}
    }
    \caption{WTMAD-2 for various methods on the selected subsets of GMTKN55.\textsuperscript{\emph{a}} Unit in kcal/mol.} 
    \begin{tablenotes}
        \item[1] \textsuperscript{\emph{a}} Best in \textbf{bold} excluding DFT.
        \item[2] \textsuperscript{\emph{b}} Overall WTMAD-2 for subsets consists of H, C, N and O, the AIQM1 collection.
        \item[3] \textsuperscript{\emph{c}} Overall WTMAD-2 for the listed subsets as a whole, see \textbf{Computational Details}.
    \end{tablenotes}
    \label{tab:gmtkn55}
\end{table}

As shown in Table \ref{tab:gmtkn55}, while the overall WTMAD-2 values are tightly contested (17.20 vs. 18.45 kcal/mol), XPaiNN outperforms GFN2-xTB in the majority of subsets. Major advantages are evident in scenarios involving energy differentials between molecules at nearly equilibrium geometries, notably in subsets such as FH51, TAUT15, DARC, ISO34, and those focusing on intramolecular NCIs. These achievements can largely be attributed to the model capacity of the XPaiNN architecture as well as the high-quality training dataset, as evidenced by the performance of the target $\omega$B97M-D3(BJ) method. Encouragingly, XPaiNN also exhibits competitiveness akin to the AIQM1 model, as validated by outcomes within the HCNO subset, the domain of proficiency of the AIQM1 model. While AIQM1 shows remarkable accuracy in predicting reaction energies as in DARC, BSR36, and ISO34, but falls short in accurately depicting NCIs as in ADIM6, IDISP, PCONF21, and SCONF. In contrast, XPaiNN attains an overall accuracy on par with AIQM1 within this HCNO subset, being better in modeling intermolecular NCIs of ADIM6, S22 and S66. As the training process embodies these NCIs within conformation energies, thereby enhancing the model’s descriptive power in this area.

Meanwhile, it becomes apparent that XPaiNN faces challenges with systems that exhibit chemical characteristics that are different from its training data. This includes large, highly conjugated systems such as fullerene (C60ISO), molecular configurations at transition states, specifically the determination of reaction barrier heights in proton transfers (PX13 and WCPT18), intermolecular NCIs mediated by halogen bonds (HAL59), and intramolecular NCIs within polypeptide chains (PCONF21). Some of these challenges are similarly encountered by AIQM1, highlighting shared limitations in handling chemically or structurally different systems as compared to their training data.

Notably, XPaiNN@xTB and OrbNet-Equi perform fairly well on these assessments with an overall improved WTMAD-2 of 10.44 and 11.77 kcal/mol, respectively. In several subsets, they approach the level of accuracy of the targeted DFT method. Of great significance, it is seen that for the intricate systems that pose challenges to direct-learning models, these $\Delta$-ML models exhibit an enhanced transferability. Notably, XPaiNN@xTB improves upon XPaiNN, achieving results that rival DFT-level accuracy as in C60ISO and WCPT18, although some advancements remain partial as in PX13 and HAL59. These findings underscore the significance of incorporating physical principles into atomistic ML models to ensure robust and transferable predictive power. While SQM methods stand out as efficient and appealing choices of the baseline methods, their accuracy in turn impacts the models' performance, as evidenced in subsets such as TAUT15 and PCONF21. All in all, the commendable accuracy of the $\Delta$-ML models on the GMTKN55 benchmark is a synergistic outcome of both model training strategies and the choice of SQM baselines.

\textbf{The barrier height benchmark for pericyclic organic reactions}. Reaction barrier height, a key factor in determining reaction kinetics, is defined as the energy difference between the reactant and the transition state (TS). A prior study\cite{aiqm1_bh} revealed limitations in the model capacity of AIQM1 to accurately predict barrier heights for pericyclic organic reactions. Consequently, we assess our models using the same dataset, which is a subset of the BH9 dataset\cite{bh9}, where reference values have been calculated utilizing the DLPNO-CCSD(T)/CBS level of theory. Our evaluation involves not only the barrier heights for both the forward and reverse reactions but also 102 reaction energies in total. The results are presented in Table \ref{tab:bh9}.

\begin{table}[htbp]
    \renewcommand{\arraystretch}{1.3}
    \centering
    \resizebox{\textwidth}{!}{
        \begin{tabular}{cccccccc}
            \toprule
            & $\omega$B97M-D3(BJ) & XPaiNN & XPaiNN@xTB & ANI-1ccx & AIQM1 & GFN2-xTB & ODM2 \\
            \midrule
            Forward Barrier & 1.13 & 24.38 & 4.93 & 8.06 & 13.24 & 12.39 & 7.51	\\
            Backward Barrier& 0.96 & 23.12 & 4.93 & 7.24 & 13.70 & \enspace4.67 & 8.74 \\
            Reaction Energy & 1.13 & \enspace2.71 & 1.36 & 3.56 & \enspace2.05 & 14.68 & 4.77 \\
            \bottomrule
        \end{tabular}
    }
    \caption{Mean absolute errors on predicting reaction barrier heights and energies on the selected reactions of BH9.\textsuperscript{\emph{a}} Unit in kcal/mol.}
    \begin{tablenotes}
        \item \textsuperscript{\emph{a}} Containing closed-shell compounds of H, C, N and O, mainly consists of Diels-Alder and other Pericyclic reactions. 
    \end{tablenotes}
    \label{tab:bh9}
\end{table}

Clearly, the accuracy of the ML models in predicting barrier heights of pericyclic reactions is not consistently reliable. Among these, XPaiNN performs the least satisfactorily. The performance of ANI-1ccx is comparable to that of SQM methods, while AIQM1 fails to outperform its baseline, ODM2. However, XPaiNN@xTB demonstrates a fairly good accuracy, with MAD of 4.93 kcal/mol for both forward and reverse barriers, marking a substantial improvement over the standalone XPaiNN, whose deviations exceed 20 kcal/mol. Reflecting on the performance of AIQM1, we attribute the enhanced transferability of XPaiNN@xTB to the synergistic effects of the $\Delta$-ML strategy and the power of equivariant GNNs. This serves as further conclusive evidence supporting our benchmark results of GMTKN55.

Noteworthily, Table \ref{tab:bh9} reveals a clear discrepancy in the performance of all ML models when comparing their predictions of reaction energies versus barrier heights. While these models are outstanding at estimating reaction energies from equilibrium structures, their performances degrade in predicting reaction barriers, which can undoubtedly be attributed to the absence of TS structures within the training dataset SPICE. This deficiency emphasizes the necessity to incorporate TS structures into the training set when developing general-purpose atomistic ML models for consistent description of a chemical reaction along the reaction pathway.

\textbf{The S66x8 and Torsion benchmarks}. We use the S66x8\cite{s66x8} and Torsion\cite{torsion} datasets as benchmarks to further evaluate the performances of XPaiNN and XPaiNN@xTB. Both datasets contain over 60 organic molecular systems, each with a diverse range of conformational profiles. Specifically, the S66x8 dataset focuses on intermolecular NCIs, containing hydrogen bonding, $\pi$-$\pi$ stacking, $X$-$\pi$ interactions, and aliphatic dispersion interactions, etc., while the Torsion dataset emphasizes intramolecular rotational conformations. Both datasets employ highly accurate CCSD(T)/CBS calculations as energy references. In order to facilitate a comparison with the ANI potentials, we concentrate on systems composed of H, C, N, and O, which results in a selection of 45 profiles from the original 62 in the Torsion dataset.

\begin{figure}[htbp]
    \centering
    \includegraphics[width=0.9\textwidth]{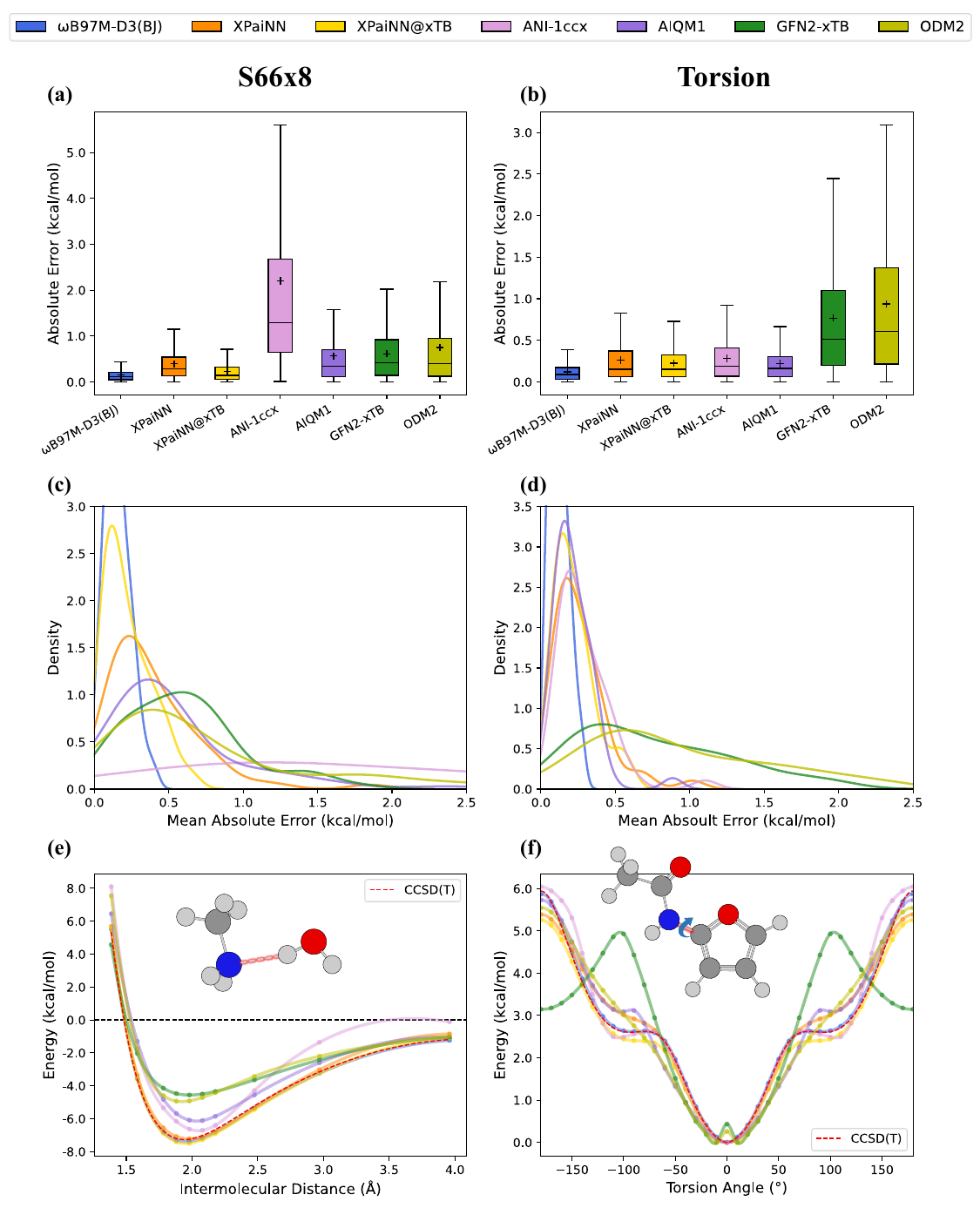}
    \caption{
        \textbf{Results of different methods on S66x8 and Torsion}. 
        \textbf{(a) (b)} Box plots for absolute error of different methods. The horizontal line is the median and the "+" symbol is the mean. 
        \textbf{(c) (d)} MAE distribution of different methods. The MAEs are Averaged by molecules or molecule pairs. 
        \textbf{(e)} Example of intermolecular interaction between methylamine and water. The potential energy curves are shown as functions of the N-H hydrogen bond length.
                     (Two additional coordinates are included, see \textbf{Computational Details}.)
        \textbf{(f)} Example of torsion conformation for N-(furan-2-yl)acetamide. The conformational curves are shown as functions of the N-C torsion angle.
    }
    \label{fig:s66_tor}
\end{figure}

The overall error performances are depicted in Figure \ref{fig:s66_tor}, while detailed statistical analyses are provided as the zip file in the \textbf{SI}. As can be seen from Figure \ref{fig:s66_tor} \textbf{(a)} to \textbf{(d)}, our models distinguish themselves in these predictions, exhibiting small MAEs and tight error distributions, surpassing the performance of the SQM methods. As is clear, our ML models perform fairly well on the Torsion dataset, given their extensive training on conformational variations using the SPICE dataset\cite{spice}. While ANI-1ccx is less accurate in dealing with intermolecular NCIs, the AIQM1 model shows a better performance in this regard. This disparity in handling intermolecular NCIs is further highlighted when comparing XPaiNN with XPaiNN@xTB. More intriguingly, as shown in Figure \ref{fig:s66_tor}\textbf{(f)}, XPaiNN@xTB accurately predicts the torsional conformations of N-(furan-2-yl)acetamide, even in the presence of unphysical barriers as predicted by GFN2-xTB, indicating the robustness of XPaiNN@xTB in delineating intricate potential energy surfaces of NCIs.

\textbf{The ROT34 benchmark}. Geometry optimization of molecules is a common practice in computational chemistry, involving an iterative procedure to find the structure that minimizing the energy. This necessitates the calculation of energy gradients with respect to the geometry, a task that can be computationally intensive when employing a first-principles method. ML models offer a means to expedite this process. Here, we validate our models, XPaiNN and XPaiNN@xTB, to undertake geometry optimization tasks on a selection of 12 organic molecules from the ROT34 dataset\cite{rot34}. The reference geometries for these molecules have been established using the SCS-MP2\cite{scs-mp2}/def2-QZVP level of theory, which is known to closely align with experimental rotational constant data\cite{rot34}, ensuring a high degree of accuracy for benchmarking purposes.

\begin{figure}[htbp]
    \centering
    \includegraphics[width=\textwidth]{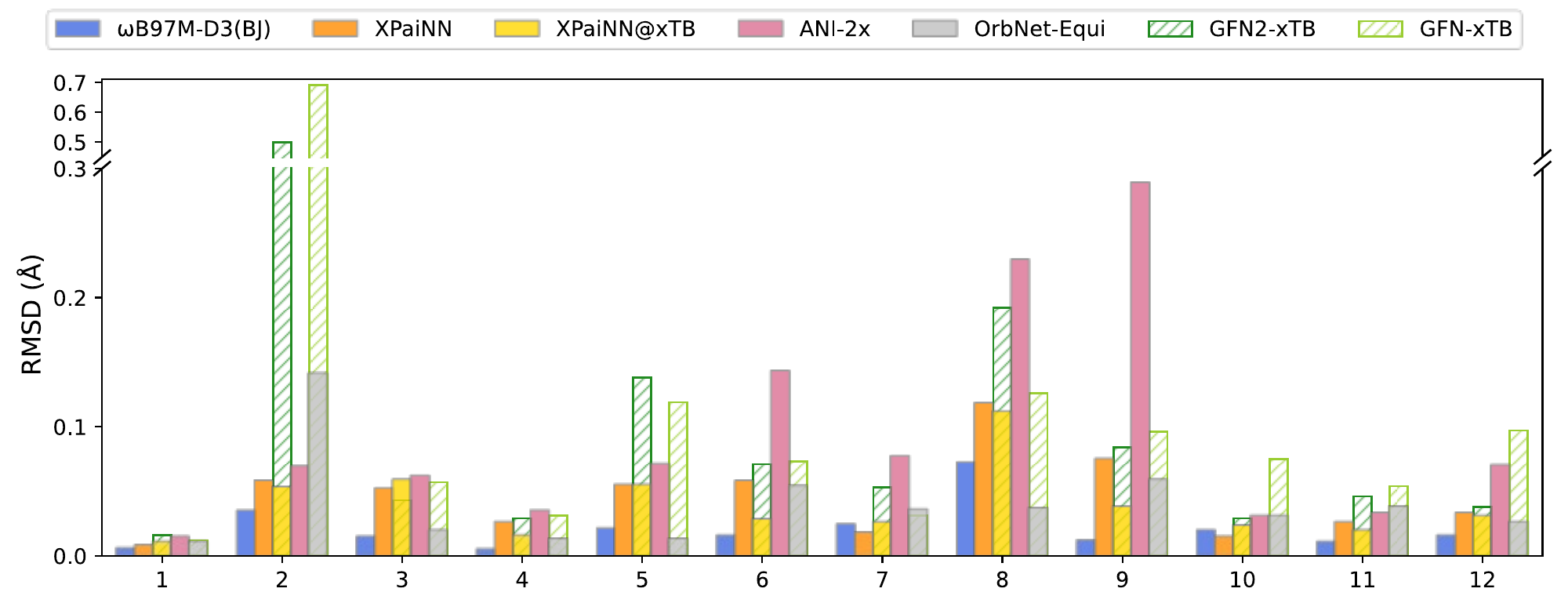}
    \caption{Root-mean-square deviations of different methods' optimized structures against reference structures for the 12 molecules in ROT34 dataset. Compound 9 is in the training dataset.}
    \label{fig:rot34}
\end{figure}

The root-mean-square deviations (RMSDs) of each optimized geometry, as determined by various methods with respect to the reference geometry, are depicted in Figure \ref{fig:rot34}. Both XPaiNN models yield superior geometric structures as compared to the SQM methods and the ANI-2x model. Specifically, they achieve average RMSDs of 0.045 \AA\, and 0.039 \AA, respectively, which are on par with that of OrbNet-Equi (0.040 \AA) and close to the target DFT accuracy (0.021 \AA). This notable performance is attributed to the comprehensive coverage of molecular conformations included in the training dataset. Consequently, our models exhibit a promising capability to perform geometry optimizations with a fairly good accuracy but at a substantially low cost.

\textbf{The reaction thermodynamics test}. We proceed to challenge our models with the estimation of reaction enthalpies, a task that is more demanding. This endeavor necessitates the calculation of vibrational frequencies at the respectively optimized molecular structure, which involves computing second-order derivatives of energy with respect to atomic positions, a process that is computationally expensive when employing sophisticated methods. Consequently, more cost-effective alternatives are frequently sought for approximating thermochemical effects\cite{bursch2022DFTProtocol}. We show here that our ML models present a viable solution by facilitating these routine calculations concurrently, offering a significantly reduced computational cost while still maintaining a satisfactory level of accuracy.

\begin{figure}[htbp]
    \centering
    \includegraphics[width=\textwidth]{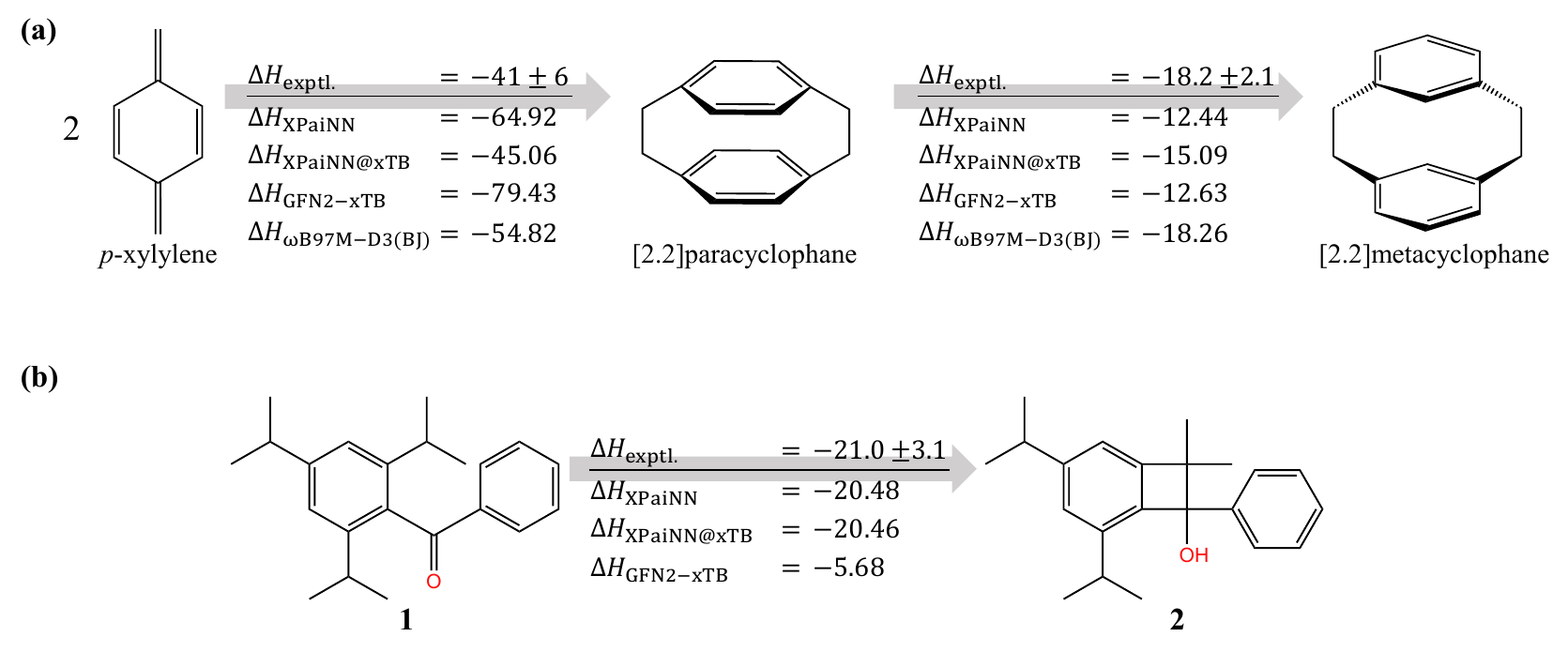}
    \caption{
        Reaction enthalpies given by different methods along with the corresponding experimental values. Unit in kcal/mol.
        \textbf{(a)} Formation of [2.2]paracyclophane from \textit{p}-xylylene and subsequent isomerization to [2.2]metacyclophane.
        \textbf{(b)} Isomerization between \textbf{1} (2,4,6-triisopropylbenzophenone) and \textbf{2} (3',5'-diisopropyl-4,4-dimethyl-3-phenyl-1,2-benzocyclobuten-3-ol).
    }
    \label{fig:thermo}
\end{figure}

Figure \ref{fig:thermo}\textbf{(a)} depicts an example for the dimerization of p-xylylene to [2.2]paracyclophane, followed by its isomerization to [2.2]metacyclophane. Here, we employ both XPaiNN and XPaiNN@xTB for geometry optimizations and subsequent vibrational frequency calculations of the involved molecules to ascertain the reaction enthalpies. For comparison, similar computations were also carried out using GFN2-xTB and the $\omega$B97M-D3(BJ) method. The results, along with experimental reference values\cite{thermo1-1, thermo1-2, thermo1-3}, are presented in Figure \ref{fig:thermo}\textbf{(a)}. Our models demonstrate enhanced accuracy in estimating reaction enthalpies compared to GFN2-xTB. Notably, the predictions from XPaiNN@xTB closely agree with the experimental values and rival those obtained from the DFT calculations. It is crucial to highlight that employing a high-level DFT approach such as $\omega$B97M-D3(BJ)/def2-TZVPPD requires 3.7 CPU-days on a machine with Intel Xeon Platinum 9282 CPU, whereas our ML models accomplish the same tasks in mere 140 CPU-seconds using one Intel Xeon Gold 6254 CPU and one NVIDIA RTX2080 GPU, thereby achieving a striking reduction in computational cost.

Figure \ref{fig:thermo}\textbf{(b)} gives another illustrative case, involving enthalpy calculations for the isomerization between compounds \textbf{1} and \textbf{2}, a process too computationally demanding to be practically undertaken using high-level DFT coupled with a large basis set. Conversely, such calculations pose no challenge to our models, and they yield results that are consistent with experimental values\cite{thermo2}. Remarkably, even though our models are trained without explicitly fitting the Hessian matrix, they manage to attain a satisfactory level of accuracy in predicting thermodynamic properties. These results serve as compelling demonstrations of the practical utility of the XPaiNN@xTB model, offering DFT-level accuracy in routine QM calculations with the computational efficiency akin to SQM methods.

\section{Discussion}

In this work, we propose XPaiNN, an equivariant GNN, designed for constructing a versatile atomistic ML model of general-purpose. The major extension from the original PaiNN to XPaiNN lies in replacing vector feature channel $\mathbf{v}$ of the former with spherical feature channel $\boldsymbol{\chi}$ in the latter, with dual objective of enhancing model capacity and data efficiency while preserving computational efficiency. The outcomes from evaluations on well-established benchmark dataset QM9 decisively endorse our design principles underlying this neural network model.

To address the challenge of model transferability, we have trained two XPaiNN models, applying both direct-learning and $\Delta$-ML strategy on the SPICE dataset. This is the first time these two training methodologies have been directly compared within the same framework, offering valuable insights into their relative performance and applicability. The direct-ML model exhibits commendable performance chemically similar systems at near-equilibrium structures, showing a competitive performance of the AIQM1 model. Meanwhile, the $\Delta$-ML model demonstrates enhanced accuracy and transferability across extensive tests, particularly for systems with diverse chemistry, including large conjugated compounds, intermolecular NCIs, and transition states. This enhancement is achieved by incorporating the SQM method GFN2-xTB as the baseline. The resultant model, XPaiNN@xTB, has been shown to offer an accurate alternative to high-level DFT methods in routine QM calculations at the speed and cost of SQM methods.

We attribute the current success to the incorporation of physical (fundamental symmetries encoded in equivariant GNNs) and chemical (systematically approximated SQM method as the baseline) priors as inductive biases during model construction. We conclude that the quality of training data and the model's architecture are equally crucial in achieving desired levels of accuracy and efficiency.

Although the current XPaiNN has not explicitly considered electronic degrees of freedom, such as charge and spin, we are actively exploring ways to integrate these effects so as to broaden the model's applicability across a wider range of chemical spaces\cite{cheng2022molecular, zubatyuk2021AIMNSE, unke2021spookynet}. Furthermore, GNN models are inherently adaptable for extended training through active learning strategies\cite{guan2023EGNNactlearning}, a beneficial trait for developing general-purpose atomistic ML models. Overall, we envision this work contributing significantly to the ongoing development of such models, offering tangible benefits to various scientific disciplines.

\section{Materials and Methods}
Model training, testing and inference with XPaiNN framework were performed using the XequiNet package (see \textbf{Code Availability}), built upon PyTorch 2.0.1\cite{pytorch}, torch-geometric 2.3.1\cite{pyg} and e3nn 0.5.1\cite{e3nn_software}. The contributions of the baseline GFN2-xTB in XPaiNN@xTB were calculated with TBlite 0.3.0\cite{tblite}. Geometry optimizations and vibrational frequence calculations for these models were carried out using geomeTRIC 1.0\cite{tric} and PySCF 2.4.0\cite{pyscf2}, respectively.

\subsection{Computational Details}
For benchmark on GMTKN55, the subset-wise error metric is define as\cite{orbnet-equi}
\begin{equation}
    \mathrm{WTMAD}\text{-}2_i = \frac{1}{N_i} \sum_{j} \mathrm{WTAD}_{i,j}
\end{equation}
\begin{equation}
    \mathrm{WTAD}_{i,j} = \frac{56.84 \thinspace \mathrm{kcal/mol}}{\frac{1}{N_i} \sum_{j} \left| \Delta E_{i,j} \right|} \cdot \left| \Delta E_{i,j} - \Delta \hat{E}_{i,j}\right|
    \label{equ:wtad}
\end{equation}
where $\mathrm{WTAD}_{i,j}$ refers to the weighted mean absolute deviation of $j$-th reaction within subset $i$, $N_i$ is the total number of reactions in subset $i$, while $\Delta E_{i,j}$ and $\Delta \hat{E}_{i,j}$ correspond to the reference and predicted reaction energies, respectively. 56.84 kcal/mol is the averaged absolute reaction energy across the entirety of the GMTKN55 dataset. It should be noted that the original definition for $\mathrm{WTAD}_{i,j}$, i.e., Equation (38) in Ref.~\citenum{orbnet-equi} is incorrect; Eq.(\ref{equ:wtad}) herein reflects the actual formula for its computation. Furthermore, the overall WTMAD-2 for both the whole dataset and the curated selection was evaluated according to
\begin{equation}
    \mathrm{WTMAD}\text{-}2 = \frac{1}{\sum_i N_i} \sum_{i,j} \mathrm{WTAD}_{i,j}
\end{equation}

The subset-wise WTMAD-2 values in Table \ref{tab:gmtkn55} for the SQM methods and other ML models were retrieved from Ref.~\citenum{orbnet-equi}. The corresponding values for $\omega$B97M-D3(BJ) were calculated using ORCA 5.0.4\cite{orca5}, while those for the AIQM1 model were attained with MLatom, version 2.0\cite{mlatom2}.

For the results pertaining to the S66x8 and Torsion benchmark datasets, calculations for GFN2-xTB were performed with xTB 6.4.0\cite{xtb}. Those for AIQM1, ANI-1ccx, and ODM2 were gathered from Ref.~\citenum{aiqm1}. For the $\omega$B97M-D3(BJ)/def2-TZVPPD calculations, ORCA 5.0.4 was utilized. For Torsion benchmark, each molecule underwent constrained optimization within its profile, with fixed dihedral angles, to determine the conformational energy. In Figure \ref{fig:s66_tor}\textbf{(e)}, two conformations at distances of 1.38 \AA\, and 1.58 {\AA} were added to the original profile, where the geometries and reference values associated with these conformations were taken from the S66x10 dataset\cite{s66x10}. The values for AIQM1 and ANI-1ccx as displayed in the figure were calculated using MLatom 2.0, while those for ODM2 were calculated using MNDO\cite{mndo}.

For the prediction of barrier heights in the BH9 dataset, the results for GFN2-xTB were computed using xTB 6.4.0. All other results in this context were sourced from Ref.~\citenum{bh9}. Regarding geometry optimizations, the optimized structures for SQM methods, ANI-2x, and OrbNet-Equi were obtained from Ref.~\citenum{aiqm1_bh}. Calculations for $\omega$B97M-D3(BJ) were performed using ORCA 5.0.4. In the case of reaction enthalpies, all reaction enthalpy values were determined using the aforementioned software programs and packages.

\subsection{Code Availablility}
The XPaiNN model is available in XequiNet package from Github: \url{https://github.com/X1X1010/XequiNet}.

\subsection{Data Availablility}
Partitioned QM9 dataset in HDF5 format, trained models on QM9, and optimized structures of ROT34 have been deposited in Zenodo (\url{https://zenodo.org/records/12745133}). Detailed and additional results on downstream benchmarks of GMTKN55, S66x8, Torsion and BH9 are provided with this work as supporting information. Requests for trained models on SPICE should be addressed to Jianming Wu and Xin Xu.

\begin{acknowledgement}
This project is supported by the National Natural Science Foundation of China (Grant 22393911, 22233002 and 21373053), and Innovation Program for Quantum Science and Technology (2021ZD0303305). The authors appreciate both the High-End Computing Center and the CFFF platform of Fudan University for generously providing computational resources.
\end{acknowledgement}

\bibliography{references}

\end{document}


\maketitle

\newpage

\section{Equivariant graph neural network}
Atomistic systems can be represented as coordinate systems in 3D Euclidean space. Transformations that preserve Euclidean distance between any two points in 3D space form the group of $E(3)$. Specifically, this group encompasses three types of transformations:  translation $\hat{T}$, rotation $\hat{R}$ and inversion $\hat{I}$. Furthermore, $\hat{T}$ and $\hat{R}$ form the group $SE(3)$. $SO(3)$ group only considers rotation symmetry, while $O(3)$ group consists of $\hat{R}$ and $\hat{I}$, which are subgroups of $E(3)$. \cite{e3nn_paper, equiformer}

The conventional strategy of using relative position vectors $\{\vec{r}_{AB}\}$ in graph representation fulfills the translation equivariance fora Graph Neural Network (GNN) model. This equivariant GNN differs from the invariant model by further fulfilling rotation and inversion ($O(3)$ group) equivariance. As a result, GNNs enbale the hidden feature and the output being transformed consistently alongside changes the input space. 
\begin{equation}
    \tilde{\mathcal{Y}} \equiv \mathcal{F} \left(\mathcal{R} \cdot \mathbf{x}\right) = \mathcal{R} \cdot \mathcal{F}\left(\mathbf{x}\right) \equiv \mathcal{R} \cdot \mathcal{Y}, \thinspace \mathcal{R} \in \{\hat{R}, \hat{I} \}
\end{equation}

Many equivariant models are $SE(3)$-equivariant\cite{unke_phisnet_2021, equiformer}, where inversion operation $\hat{I}$ is ignored. This is achieved via operating on geometric tensors $\bm{\chi}$ in spherical representation consists of direct sum of irreducible representations (Irrep) $\bm{\omega}^{(l)}$ of $SO(3)$ group  by different angular momentum degree $l$.
Models operating on geometric tensors in Cartesian form also exhibit SE(3)-equivariance. Two examples PaiNN\cite{painn} and TensorNet\cite{tensornet}. PaiNN adopts separate scalar $\bm{s} (l=0)$ and vector $\mathbf{v} (l=1)$ feature channels, while TensorNet decomposes rank-2 tesnor representation into diagnol ($l=0$), off-diagnol ($l=1$) and trace-less ($l=2$) part. These separations or decompositions allow the model to handle different types of interactions and symmetries effectively.

In the current work, the XPaiNN architecture is also $SE(3)$-equivariant.

\section{Model Details}
\subsection{Detailed Architecture}
\textbf{Node Embedding}. A well-recognized weakness of the ANI potentials\cite{ani} is their inability to predict properties of chemical compounds  that contain elements not present in the dataset used for model training. Thus, introducing a new type of element requires a bottom-up reconstruction of the model\cite{ani-2x}. GNNs avoid this defect via element-wise node embedding. However, the widely used embedding for atoms in GNN is one-hot encoding, which cannot ensure reasonable inference results for the aforementioned situations, as it fails to reflect the chemical relations as dictated by the periodic table law of elements. By contrast, OrbNet\cite{qiao2020orbnet, orbnet-equi} utilizes diagonal blocks of matrices from quantum mechanics (QM) calculations as node embedding feature, which empowers the model to adeptly manage compounds composed of untrained elements. 

Drawing inspiration from the strengths of OrbNet and recognizing that its superiority partly stems from the application of specific basis sets in QM calculations, we come up with the design of the node embedding scalar feature by projecting element-wise atomic orbital basis sets onto a uniform set of auxiliary basis. By doing so, we create features that  vary periodically, akin to the element properties across the periodic table,, while also maintaining a consistent dimensionality. This ensures that our features can be uniformly applied and compared, regardless of the specific element under consideration, thereby enriching the representational capacity of our model.

\begin{figure}[htbp]
    \centering
    \includegraphics[width=0.9\textwidth]{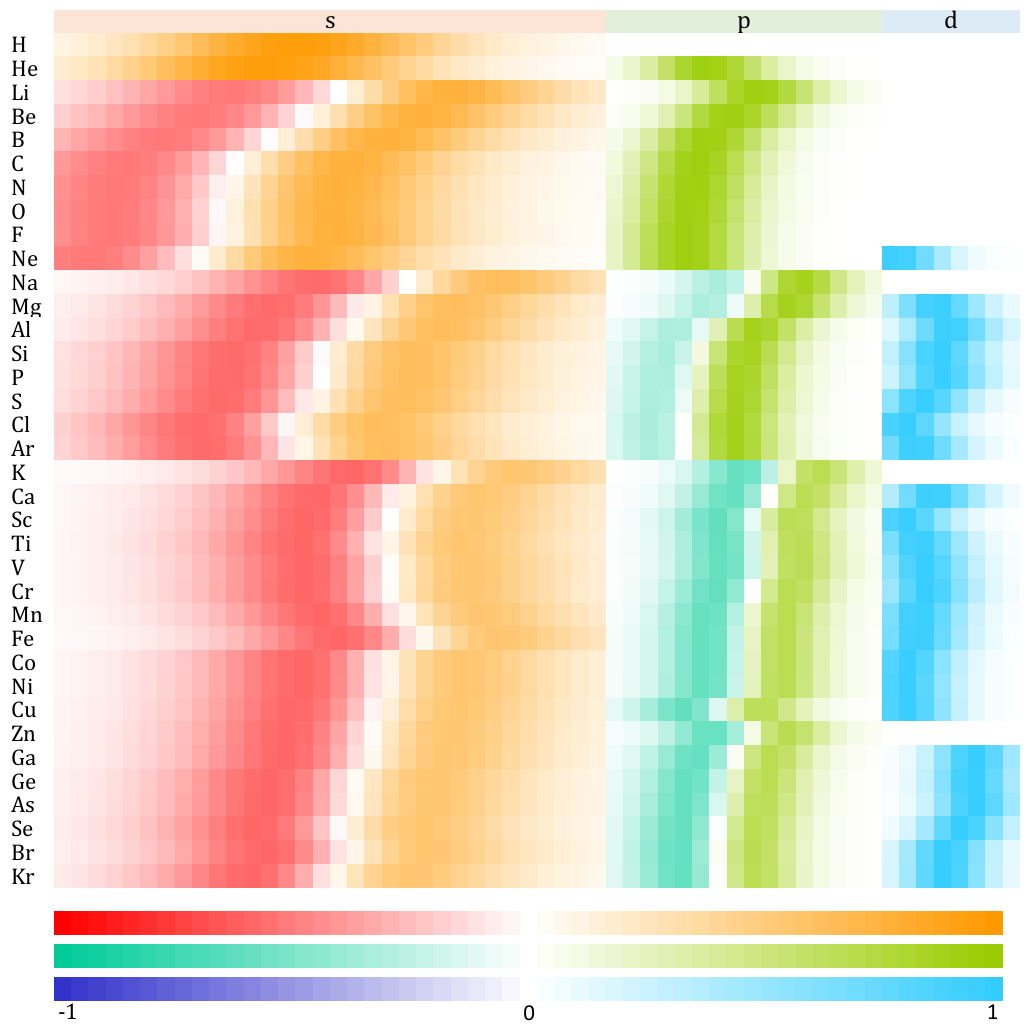}
    \caption{Visualization of atom embeddings. Color shades indicate embedding vector values.}
    \label{fig:vis_embed}
\end{figure}

We first define a set of Gaussian-type auxiliary basis functions $\widetilde{\Phi}_{n,l,m}$:
\begin{equation}
    \tilde{\Phi}_{n,l,m} \left( \mathbf{r} \right)
        := A_{n,l} \cdot
            r^{l} \exp \left( -\alpha_{n,l} r^{2} \right)
            \mathcal{Y}_{l,m} \left( \hat{\mathbf{r}} \right)
    \label{equ:aux_basis}
\end{equation}
where $r = \left\| \mathbf{r} \right\|$, $\hat{\mathbf{r}} = \mathbf{r} / r$, $\mathcal{Y}_{l,m} \left( \cdot \right)$ is the real spherical harmonics for given angular and magnetic quantum numbers $l$ and $m$, and $A_{n,l}$ is the normalization factor where $n$ is the principle quantum number. In Eq. (\ref{equ:aux_basis}), the scaling parameters $\alpha_{n,l}$ are pre-determined for different $n$ and $l$ as:
\begin{equation}
    \begin{cases}
        \alpha_{n,0} := 128 \cdot \left(\frac{1}{\sqrt{2}}\right)^{n-1} & \text{where } n \in \left\{ 1,2,...,32 \right\} \\
        \alpha_{n,1} := 64 \cdot \left(\frac{1}{2}\right)^{n-1} & \text{where } n \in \left\{ 1,2,...,16 \right\} \\
        \alpha_{n,2} := 4 \cdot \left(\frac{1}{2}\right)^{n-1} & \text{where } n \in \left\{ 1,2,...,8 \right\}
    \end{cases}
\end{equation}
Afterwards, we employ the valence basis sets of GFN2-xTB\cite{gfn2-xtb} as the atomic orbital basis function $\Psi^{Z_A}_{n,l,m}$ with each element type $A$ described by the nuclear charge number $Z_A$. Then, element-wise features are computed as the overlap integrals between the two sets of orbitals at the same coordinate origin. 
\begin{equation}
    \left( \mathbf{S}^{Z_A} \right)_{\mu \nu} 
        = S^{Z_A}_{(n_1,l_1,0)(n_2,l_2,0)}
        = \left< \widetilde{\Phi}_{n_1,l_1,0} \bigg| \Psi^{Z_A}_{n_2,l_2,0} \right>
    \label{eqs:overlap}
\end{equation}
Note in Eq.(\ref{eqs:overlap}), we only calculate overlap integrals between orbitals with $m = 0$, while there is no need to calculate overlap integrals between orbitals of different $m$ due to orthogonality. Finally, the embedding feature $\mathbf{e}^{Z_A}$ is obtained by sum over $\mathbf{S}^{Z_A}$ along the atomic basis dimension $\nu$
\begin{equation}
    \left( \mathbf{e}^{Z_A} \right)_\mu = \sum_{\nu} \left( \mathbf{S}^{Z_A} \right)_{\mu \nu}
\end{equation}
The above process generates $\mathbf{e}^{Z_A}$ as an array of 56 dimensions for each element type $A$, which are visualized using the color scale as shown in Figure \ref{fig:vis_embed}. Periodic changes in the atomic orbital basis result in the periodic changes in atom embedding vectors, which reflect the periodic table law of the elements to some extent.

\textbf{Radial Basis Function (RBF)}. The interatomic distance $r_{ij}$ is represented through an expansion using the $0$-order spherical Bessel function of the first kind, a method same as that in DimeNet\cite{dimenet}. A number of RBFs are used to perform a filter-like action on the interatomic distances, with the k-th function defined as 
\begin{equation}
    R_k\left( r_{ij} \right) = \sqrt{\frac{2}{c}}
        \frac{\sin \left( f_k r_{ij} \right) }{r_{ij}}
\end{equation}
where $c$ is the cutoff radius and $f_k$ is a trainable parameter initialized with $\frac{k\pi}{c}$. These RBFs are then linearly combined and smoothly truncated with cosine cutoff function.
\begin{equation}
    \mathcal{W}_{ij} = \left[ \sum_{k=1}^n w_k R_k\left( r_{ij} \right) + b \right]
                 \cdot \frac{1}{2} \left[ \cos \left( \pi \frac{r}{c} \right) + 1 \right]
\end{equation}

\textbf{Real Spherical Harmonics (RSH)}. The relative orientations $\hat{r}_{ij}$ between atoms are expanded using RSHs. The orientation information is then introduced into the network, forming the equivariant spherical tensors as:
\begin{equation}
    \mathcal{Y}_{l,m} \left( \hat{r}_{ij} \right) =
        \begin{cases}
            \frac{1}{\sqrt{2}} \left[
                Y_{l,-m} \left( \hat{r}_{ij} \right) + \left( -1 \right)^m Y_{l,m} \left( \hat{r}_{ij} \right)
            \right]   & \text{ if } m > 0 \\
            Y_{l,0} \left( \hat{r}_{ij} \right)  & \text{ if } m = 0 \\
            \frac{1}{\sqrt{2i}} \left[
                Y_{l,m} \left( \hat{r}_{ij} \right) - \left( -1 \right)^m Y_{l,-m} \left( \hat{r}_{ij} \right)
            \right]   & \text{ if } m < 0
        \end{cases}
\end{equation}
\begin{equation}
    Y_{l,m} \left( \hat{r}_{ij} \right) = Y_{l,m} \left( \theta, \phi \right) = 
        \sqrt{
            \frac{2l+1}{4\pi} \frac{\left( l-m \right)!}{\left( l+m \right)!}
        } 
        P_l^m \left( \cos \theta \right) e^{im\phi }
\end{equation}
 $Y_{l,m}\left(\cdot\right)$ is the complex spherical harmonics and $\mathcal{Y}_{l,m}\left(\cdot\right)$ is the desired RSHs.

\begin{figure}[htbp]
    \centering
    \includegraphics[width=0.7\textwidth]{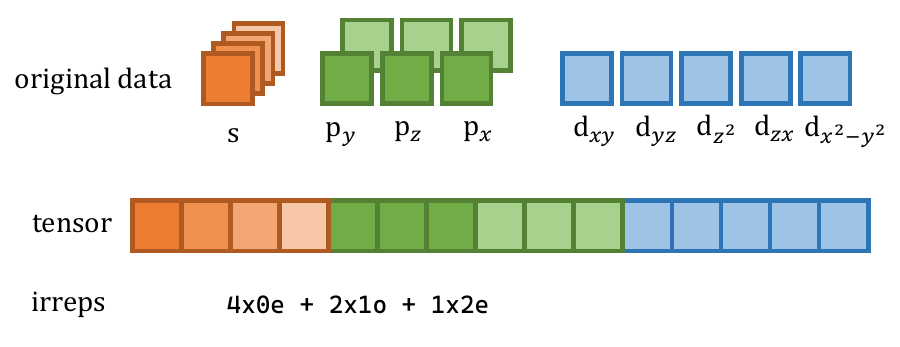}
    \caption{Schematic of Spherical Tensor}
    \label{fig:sphten}
\end{figure}

\textbf{Spherical Tensor}. Spherical Tensor serves as the basic data structure for $E(3)$ equivariant operations on tensor data. It is characterized by a direct sum of inequivalent irreducible representations of $E(3)$ group (Irreps), combined with numbers indicating feature channels in each tensor of $l$ order\cite{e3nn_paper}. The schematic is illustrated in Figure \ref{fig:sphten}.

\begin{figure}[htbp]
    \centering
    \includegraphics[width=0.6\textwidth]{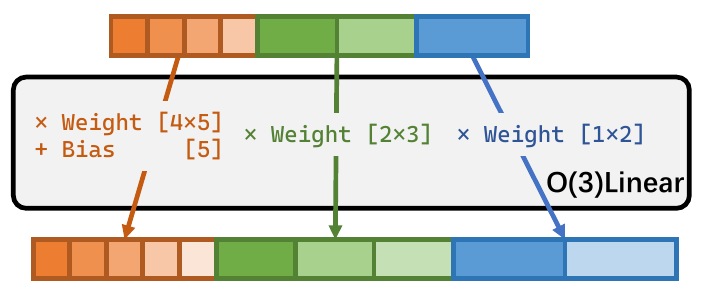}
    \caption{Schematic of $O(3)$ linear transform.}
    \label{fig:o3linear}
\end{figure}

$\bm{O(3)}$ \textbf{Linear}. Linear transformations are separately carried out for each tensor of $l$ order $\bm{\omega}^{(l,p)}$ in Irreps to ensure equivariance. In this framework, the number of channels is adjustable. Notably, the bias term can only be incorporated into the transformation for the $0$th order Irrep (i.e. the scalar-type) for $l = 0$, 
\begin{equation}
    \left[ O(3)\text{Linear}\left( \bm{\chi} \right) \right]_{l,p}
    = \begin{cases}
        \bm{\omega}^{(l,p)} \mathbf{W}_{l,p}^\top + \mathbf{b}_p & l=0 \\
        \bm{\omega}^{(l,p)} \mathbf{W}_{l,p}^\top                & l>0
    \end{cases}
\end{equation}
where $p$ denotes the parity, which refers to the inversion symmetry. The schematic is shown in Figure \ref{fig:o3linear}.

\begin{figure}[htbp]
    \centering
    \includegraphics[width=0.7\textwidth]{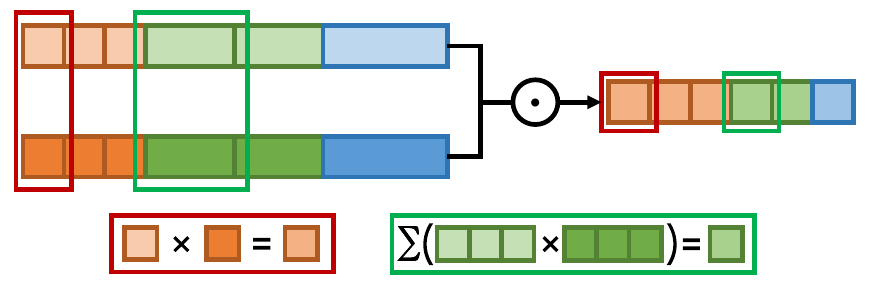}
    \caption{Schematic of inner product between two spherical tensors.}
    \label{fig:dot}
\end{figure}

$\bm{O(3)}$ \textbf{Inner Product}. To compute the inner product of two spherical tensors of the same layout, one performs element-wise multiplication and sum over the degenerate dimension, represented by $m$, for each distinct value of $l$. 
\begin{equation}
    \bm{\chi}_1 \odot \bm{\chi}_2 = \bigoplus_{l,p} \bm{\omega}_1^{(l,p)} \cdot \bm{\omega}_2^{(l,p)}
\end{equation}
\begin{equation}
    \bm{\omega}_1 \cdot \bm{\omega}_2
        = \sum_{m=-l}^{l} \left( \bm{\omega}_1 \right)_{m} \left( \bm{\omega}_2 \right)_{m}
    \label{equ:innerdot}
\end{equation}
For simplification, $(l,p)$ is omitted for $\bm{\omega}$ in Eq. \ref{equ:innerdot}. The result is a scalar tensor, as shown in Figure \ref{fig:dot}

\begin{figure}[htbp]
    \centering
    \includegraphics[width=0.7\textwidth]{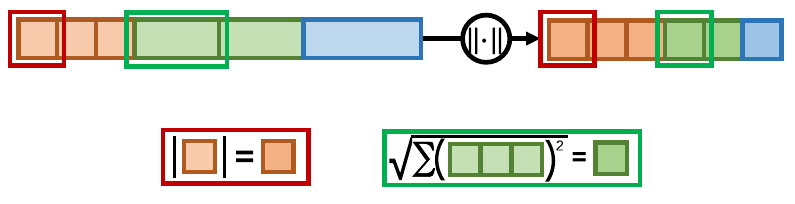}
    \caption{Schematic of invariant operation on a spherical tensor.}
    \label{fig:invariant}
\end{figure}

\textbf{Invariant}. Similar to $\bm{O(3)}$ \textbf{Inner Product}, one computes the L2-norm of each $\bm{\omega}^{(l,p)}$ as shown in Figure \ref{fig:invariant} . The result is also a scalar tensor.
\begin{equation}
    \left\| \bm{\chi} \right\| 
        = \bigoplus_{l,p} \sqrt{
            \sum_{m=-l}^l \left| \left( \bm{\omega}_m \right) \right|^2
        }
\end{equation}

\begin{figure}[htbp]
    \centering
    \includegraphics[width=0.7\textwidth]{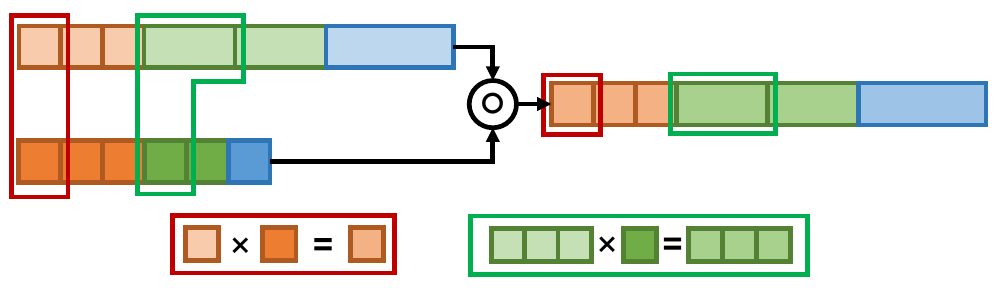}
    \caption{Schematic of Hadamard product between a spherical tensor and a scalar tensor.}
    \label{fig:hadamard}
\end{figure}

\textbf{Hadamard Product}. A Hadamard product can be computed between a spherical tensor and a scalar tensor, provided that the length of the scalar tensor is the same as the number of Irreps channel in the spherical tensor, as illustrated in Figure \ref{fig:hadamard}. In this context, $\bm{\omega}^{(l,p)}$ is multiplied, term by term, by the value of the corresponding scalar tensor.
\begin{equation}
    \left( \bm{\chi} \circ \mathbf{x} \right)_{l,m,p}
        = \left( \bm{\chi} \right)_{l,m,p} \left( \mathbf{x} \right)_{l,p}
\end{equation}

\begin{figure}[htbp]
    \centering
    \includegraphics[width=0.9\textwidth]{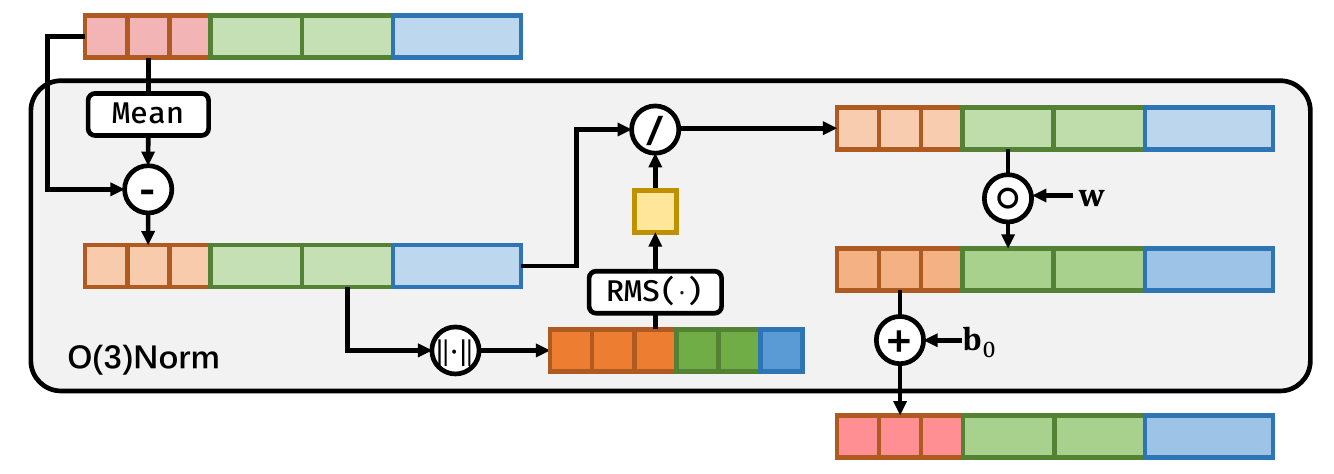}
    \caption{Schematic of the layer normalization of a spherical tensor.}
    \label{fig:norm}
\end{figure}

$\bm{O(3)}$ \textbf{Normalization}. The equivariant normalization is done by deducting the mean for scalar sub-tensor, and dividing the overall spherical tensor by the root mean square of its invariant tensor as shown in Figure \ref{fig:norm}.
Note, the "$+$" and "$-$" is performed only for the $0$th order Irreps. $\mathbf{w}$ and $\mathbf{b}_0$ are parameters to be optimized. $\text{RMS}\left(\cdot\right)$ refers to root mean square.

\begin{figure}[htbp]
    \centering
    \includegraphics[width=0.9\textwidth]{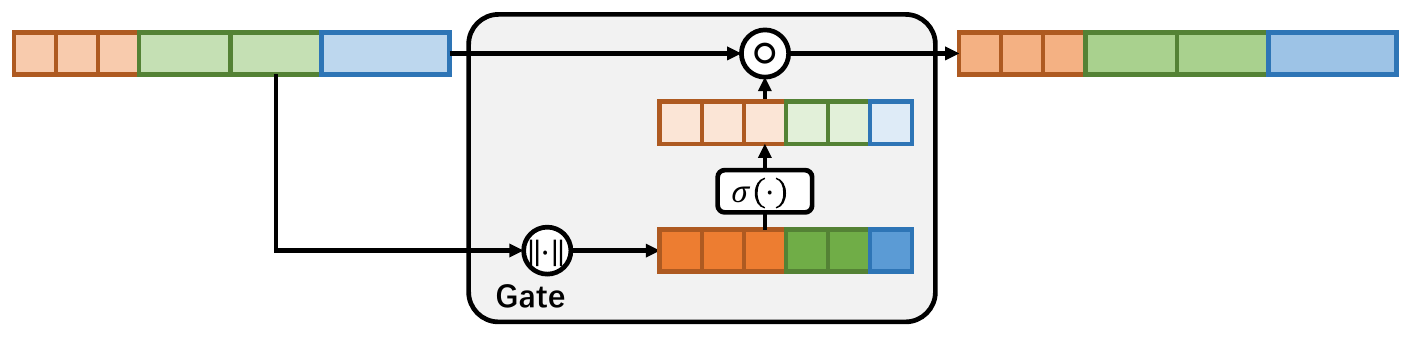}
    \caption{Schematic of activation gate of a spherical tensor.}
    \label{fig:gate}
\end{figure}

\textbf{Activation Gate}. The activation gate for spherical tensor is constructed based on the sigmoid linear unit (SiLU) activation function as shown in Figure \ref{fig:gate}.
\begin{equation}
    \text{SiLU}\left( x \right) = x * \sigma\left( x \right)
\end{equation}
\begin{equation}
    \text{Gate}\left( \chi \right) = \bm{\chi} \circ \sigma \left(
        \left\| \bm{\chi} \right \|
    \right)
\end{equation}
Here, $\sigma\left(\cdot\right)$ is the sigmoid function, i.e. $\sigma\left(x\right)=\left(1+e^{-x}\right)^{-1}$.

\subsection{Node updating}
\begin{figure}
    \centering
    \includegraphics[width=0.5\textwidth]{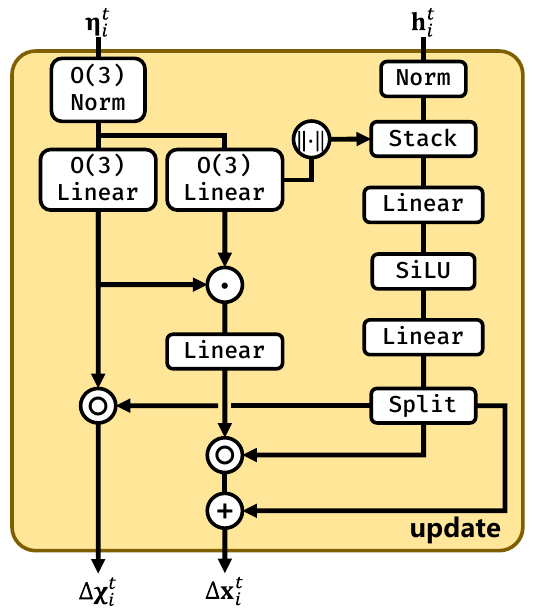}
    \caption{Schematic of node updating layer.}
    \label{fig:update}
\end{figure}

In the node updating layer depicted in Fig. \ref{fig:update}, scalar and spherical features are coupled to update both, while the operations on spherical features are localized within each irreducible representation to ensure the equivariance.
\begin{equation}
    \mathbf{x}_i^t = \mathbf{h}_i^t + 
        f_{u}^t\left( \mathbf{h}_i^{t}, \bm{\eta}_i^{t}\right)
\end{equation}
\begin{equation}
    \bm{\chi}_i^t = \bm{\eta}_i^t + 
        \phi_{u}^t\left( \mathbf{h}_i^{t}, \bm{\eta}_i^{t}\right)
\end{equation}

\begin{figure}[htbp]
    \centering
    \includegraphics[width=0.85\textwidth]{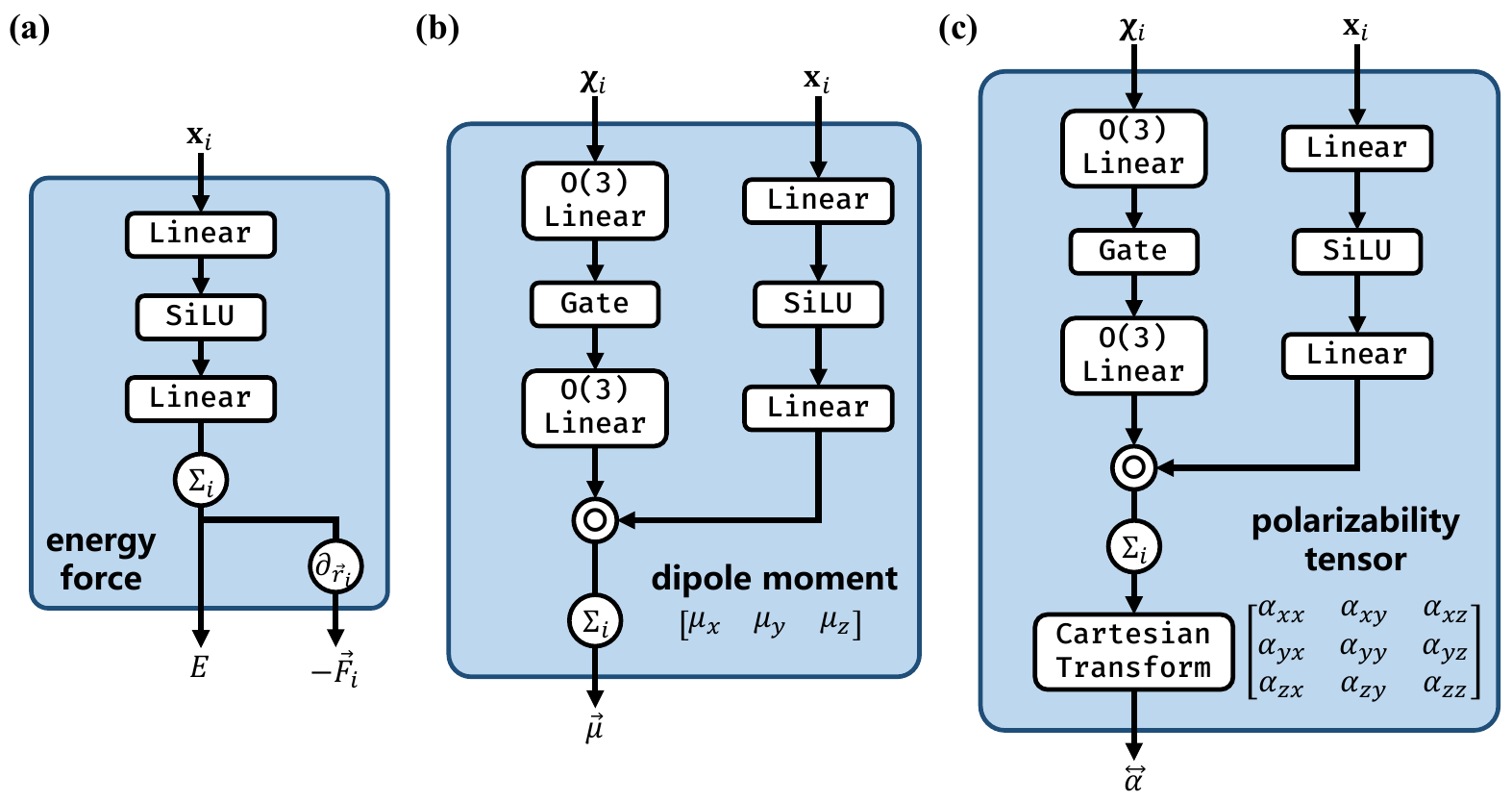}
    \caption{
        \textbf{Schematic diagram of output layers. Tensor dimensions and irreducible representations are shown in grey.} 
        \textbf{(a)} Energy and gradient force output. 
        \textbf{(b)} Dipole moment output. 
        \textbf{(c)} Polarizability tensor output.
    }
    \label{fig:output}
\end{figure}

\subsection{Output}

The final output layer are designed according to the target property.

\textbf{Energy}, a scalar property, can be constructed through scalar features (Figure \ref{fig:output}\textbf{(a)}). \textbf{Force}, an atom-wise feature, is obtained by taking derivative of the energy with respect to the coordinates.
\begin{equation}
    E = \sum_i f_o^{E} \left( \mathbf{x}_i \right),
    \quad
    \vec{F}_A = -\frac{\partial E}{\partial \mathbf{r}_A}
\end{equation}
\textbf{Dipole moment}, a vector feature, is derived from the p-type part of spherical features (Figure \ref{fig:output}\textbf{(b)})
\begin{equation}
    \vec{\mu}
        = \left( \mu_x, \mu_y, \mu_z \right)
        = \sum_i \phi_o^{\mu} \left( \mathbf{x}_i, \bm{\chi}_i \right)
\end{equation}
The modulus of dipole moment is calculated as
\begin{equation}
    \mu = \left\| \vec{\mu} \right\| = \sqrt{\mu_x^2 + \mu_y^2 + \mu_z^2}
\end{equation}
\textbf{Polarizability tensor}, a symmetric order-2 tensor, has six degrees of freedom. Therefore, the output in spherical representation is \textsf{1x0e+1x2e}. (Figure \ref{fig:output}\textbf{(c)}).
\begin{equation}
   \left(\alpha_0, \alpha_{xy}, \alpha_{yz}, \alpha_{z^2}, \alpha_{zx}, \alpha_{x^2-y^2}\right)
   = \sum_i \phi_o^{\alpha} \left( \mathbf{x}_i, \bm{\chi}_i \right)
\end{equation}
In the output layer, the model's hidden features are first transformed into the required spherical representation. Subsequently, features of the second-order spherical harmonic type are further converted into the Cartesian form.
\begin{equation}
    \begin{split}
         \left| \alpha \right| &= \sqrt{\alpha_{xy}^2 + \alpha_{yz}^2 + \alpha_{z^2}^2 + \alpha_{zx}^2 + \alpha_{x^2-y^2}^2} \\
         \alpha_{xx} &= \frac{1}{\sqrt{3}} \left( \left|\alpha\right| - \alpha_{z^2} \right) + \alpha_{x^2-y^2} \\
         \alpha_{yy} &= \frac{1}{\sqrt{3}} \left( \left|\alpha\right| - \alpha_{z^2} \right) - \alpha_{x^2-y^2} \\
         \alpha_{zz} &= \frac{1}{\sqrt{3}} \left( \left|\alpha\right| + 2\alpha_{z^2} \right) \\
         \alpha_{xy} &= \alpha_{yx} \quad
         \alpha_{yz} = \alpha_{zy} \quad
         \alpha_{zx} = \alpha_{xz}
    \end{split}
\end{equation}
The polarizability tensor is then written in the form of a $3\times3$ matrix and $\alpha_0$ is added to each diagonal element.
\begin{equation}
    \overleftrightarrow{\alpha} = \left(
        \begin{array}{ccc}
            \alpha_{xx} & \alpha_{xy} & \alpha_{xz} \\
            \alpha_{yx} & \alpha_{yy} & \alpha_{yz} \\
            \alpha_{yz} & \alpha_{zy} & \alpha_{zz}
        \end{array}
    \right) + \left(
        \begin{array}{ccc}
            \alpha_0 & 0 & 0 \\
            0 & \alpha_0 & 0 \\
            0 & 0 & \alpha_0
        \end{array}
    \right)
\end{equation}
The isotropic polarizability is obtained by calculating the trace of $\overleftrightarrow{\alpha}$
\begin{equation}
    \alpha = \frac{1}{3}\text{tr}\left(\overleftrightarrow{\alpha}\right)
\end{equation}
\textbf{Electronic spatial extent}, for a mass-centered molecule, can be obtained from the following equation.
\begin{equation}
    \left< R^2 \right> = \sum_i 
        \left\| \mathbf{r}_i \right\|^2
        f_o^{R}\left( \mathbf{x}_i \right)
    \label{equ:r2}
\end{equation}

\newpage
\subsection{Architecture Comparison}

\begin{figure}[htbp]
    \centering
    \includegraphics[width=0.8\textwidth]{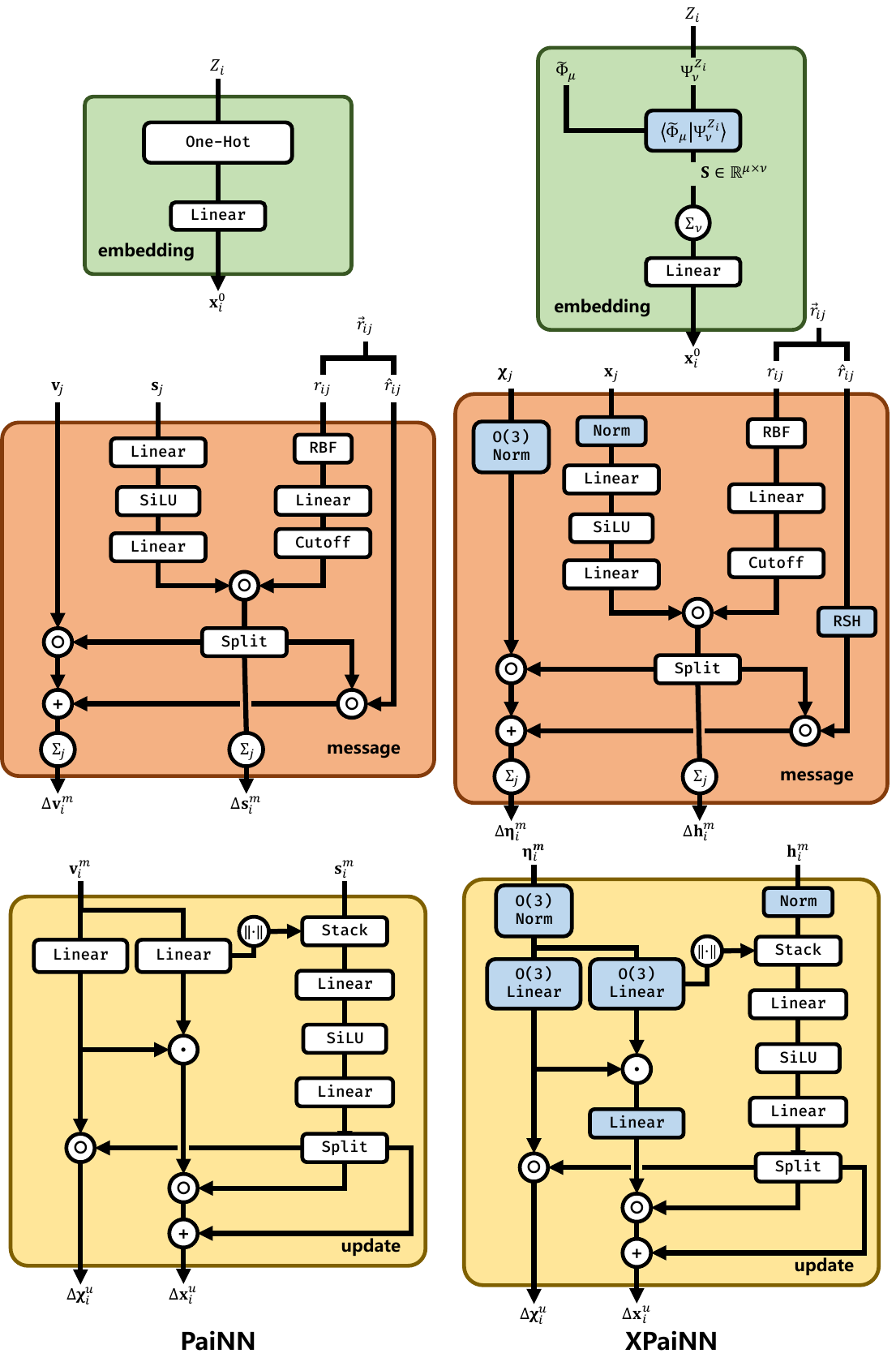}
    \caption{Comparison of the model architecture between original PaiNN and XPaiNN. The modification parts are marked as blue blocks.}
    \label{fig:difference}
\end{figure}

The main differences between XPaiNN and the original PaiNN lie in the embedding type and the form of equivariant features, which are illustrated in Fig \ref{fig:difference}. As for the embedding type, XPaiNN uses a method of orbital projection as described in \textbf{Detailed Architecture}, whereas PaiNN simply uses one-hot encoding and linear transformation to derive the initial invariant features. As for equivariant features, XPaiNN expands the relative positions of atom pairs with real spherical harmonics to obtain Irreps of different angular momentum to form equivariant features, while PaiNN only uses unit vectors of relative positions as equivariant features, which is literally the Irrep with angular momentum of $l = 1$. As a result, the corresponding layers in XPaiNN are adapted to accommodate the $O(3)$ group. In addition, XPaiNN incorporates normalization layers, which enhances the gradient flow and training stability.

\newpage 
\section{Training Details}
\textbf{QM9}. For all 12 tasks in QM9, we employ the same partition strategy, which randomly divide the dataset into three parts (110k, 10k and 11k) for training, validation and testing, respectively. For each distinct task, individual models are trained with the same hyper-parameters summarized in Table \ref{tab:hyper-param:qm9}, with the exception of the \textsf{Output hidden Irreps} parameters, which are tailored to match the target properties as depicted in \textbf{Output}. For $\left<R^2\right>$ task, we use the same output as PaiNN\cite{painn}, calculated according to Eq. \ref{equ:r2}. For certain tasks, a linear fitting based on element type towards the training set labels is performed. The as-obtained results are then subtracted from the training labels, which are subsequently reinstated during testing to refine predictions. 
For $\Delta$-learning tasks of $U_0$, $U$, $H$ and $G$, we subtract the energies of each isolated atom calculated at the respective label and baseline methods to ensure the consistency. Training for each property in QM9 was conducted using a single NVIDIA RTX2080 Ti graphics processing unit (GPU) card.

\begin{table}[htbp]
    \centering
    \begin{tabular}{ll}
        \toprule
        Hyper-parameters & Value or description \\
        \midrule
        Scalar feature dimension & 128 \\
        Spherical feature Irreps & \textsf{128x0e+64x1o+32x2e} \\
        Output hidden dimension & 64 \\
        Output hidden Irreps & \textsf{32x1o} for $\mu$, \textsf{64x0e+16x2e} for $\alpha$ \\
        Number of radius basis & 20 \\
        Cutoff radius (\AA) & 5.0 \\
        Layer normalization & False \\
        Number of interaction blocks & 3 \\
        Batch size & 100 \\
        Training epochs & 2500 \\
        Learning rate warmup & Linear warmup for 10 epochs \\
        Maximum learning rate & $5 \times 10^{-4}$ \\
        Learning rate scheduler & Cosine annealing with Tmax=500 epochs\\
        Loss function & Smooth L1 \\
        Optimizer & Adam \\
        Exponential moving average decay & 0.995 \\
        \midrule
        Total number of parameters & 861k \\
        \bottomrule
    \end{tabular}
    \caption{Hyper-parameters for training QM9 tasks.}
    \label{tab:hyper-param:qm9}
\end{table}

\newpage 
\textbf{MD17 and rMD17}. For training and validation, we randomly select 950 and 50 different configurations, respectively, and we use the remaining configurations for test. Model hyper-parameters are summarized in Table \ref{tab:hyper-param:md17}. We substract the energies of isolated atoms at the same level of the label for training, and we use the averaged energy as the initial bias for the linear layer in the output module. Training for each molecule is conducted on a single NVIDIA RTX2080 Ti GPU card.

\begin{table}[htbp]
    \centering
    \begin{tabular}{ll}
        \toprule
        Hyper-parameters & Value or description \\
        \midrule
        Scalar feature dimension & 128 \\
        Spherical feature Irreps & \textsf{128x0e+64x1o+32x2e} \\
        Output hidden dimension & 64 \\
        Number of raidus basis & 20 \\
        Cutoff radius (\AA) & 5.0 \\
        Layer normalization & False \\
        Number of interaction blocks & 3 \\
        Batch size & 10 \\
        Training epochs & 2500 \\
        Learning rate warmup & Linear warmup for 10 epochs \\
        Maximum learning rate & $5 \times 10^{-4}$ \\
        Learning rate scheduler & Cosine annealing with Tmax=500 epochs\\
        Loss function & MSE Loss \\
        Loss ratio of energy and force & 1:99 \\
        Optimizer & Adam \\
        Exponential moving average decay & 0.995 \\
        \midrule
        Total number of parameters & 861k \\
        \bottomrule
    \end{tabular}
    \caption{Hyper-parameters for MD17 and rMD17 dataset.}
    \label{tab:hyper-param:md17}
\end{table}

\newpage
\textbf{SPICE}. The training set used for constructing the general models is selected from 3 subsets in SPICE, namely PubChem, DES370K Monomers and DES370K Dimers, covering elements of H, C, N, O, F, P, S, Cl, Br and I. Systems that are not neutral or closed-shell are excluded from our selections. This resultes in a training set of nearly 1 million conformations, spanning over 18,000 organic and inorangic molecules. 

\begin{table}[htbp]
    \centering
    \begin{tabular}{ll}
        \toprule
        Hyper-parameters & Value or description \\
        \midrule
        Scalar feature dimension & 128 \\
        Spherical feature Irreps & \textsf{128x0e+64x1o+32x2e} \\
        Output hidden dimension & 64 \\
        Number of raidus basis & 20 \\
        Cutoff radius (\AA) & 5.0 \\
        Layer normalization & True \\
        Number of interaction blocks & 3 \\
        Batch size & 512 \\
        Learning rate warmup & Linear warmup for 10 epochs \\
        Maximum learning rate & $5 \times 10^{-4}$ \\
        Learning rate scheduler & Cosine annealing with Tmax=250 epochs \\
        Loss function & Smooth L1 \\
        Loss ratio of energy and force & 1:9 \\
        Optimizer & Adam \\
        Exponential moving average decay & 0.995 \\
        \midrule
        Total number of parameters & 865k \\
        \bottomrule
    \end{tabular}
    \caption{Hyper-parameters for the selected SPICE dataset.}
    \label{tab:hyper-param:spice}
\end{table}

We split the selected SPICE dataset into the training and validation sets using a 9:1 ratio for conformations in each molecule or system. Hyper-parameters are listed in Table \ref{tab:hyper-param:spice}. The models are trained parallelly on 8 NVIDIA A800 GPU cards. Training processes are carried out in a continuous way, adopting a cosine annealing learning rate scheduler until the error metrics on validation set stabilize. Upon convergence, the MAEs for energy and force settle at approximately 20 meV and 30 meV/\AA, respectively.

\newpage 
\textbf{Ablation experiment}
Ablation experiments are performed on the $U_0$ tag of QM9. We used the same training strategy described above in \textbf{QM9}. To ensure consistency in training conditions, we utilize the same dataset partitioning. Meanwhile, we have used our own implementation of PaiNN, resulting in a deviation from the initial reported MAE (5.85 eV)\cite{painn}, although this discrepancy shall not affect the comparability of ablation experiments. The results summarized in Table \ref{tab:ablation} indicate that both projection embedding and spherical harmonic expansion can improve the orignal performance result to some extent, while the latter is more beneficial. All training and testing in ablation experiments are carried out on a single NVIDIA RTX2080 Ti GPU card.

\begin{table}[htbp]
    \centering
    \resizebox{\textwidth}{!}{
        \begin{tabular}{lll}
            \toprule
            Model & MAE (meV) & Description \\
            \midrule
            PaiNN & 6.27 & Original PaiNN \\
            PaiNN + Projection embedding & 6.21 & PaiNN with auxiliary basis projection embedding \\
            PaiNN + Spherical tensor & 5.63 & PaiNN with spherical tensor \\
            XPaiNN & 5.44 & XPaiNN, including projection embedding and spherical tensor \\
            \bottomrule
        \end{tabular}
    }
    \caption{Ablation experiments on the $U_0$ tag of QM9.}
    \label{tab:ablation}
\end{table}

\newpage 
\section{Additional Results}

\textbf{$\Delta$-ML results on QM9}. An additional experiment has been performed, applying the $\Delta$-ML approach using XPaiNN@xTB to four thermodynamic targets in QM9. The results are compared to those SOTA models as listed in Table \ref{tab:qm9-delta}. As is clear, the performance of XPaiNN@xTB closely matches that of the top direct-learning models, such as TensorNet\cite{tensornet} and Allegro\cite{allegro}. Meanwhile, OrbNet-Equi\cite{orbnet-equi}(another $\Delta$-ML model) is the best of all, which largely owes to its feature generation using quantum mechanical information as physical priors. These results suggest the promising potential of $\Delta$-ML models.

\begin{table}[htbp]
    \centering
    \resizebox{0.75\textwidth}{!}{
        \begin{tabular}{llcccc}
        \toprule
        Task & Unit & Allegro\textsuperscript{\emph{a}} & TensorNet\textsuperscript{\emph{a}} & OrbNet-Equi\textsuperscript{\emph{a}} & XPaiNN@xTB\textsuperscript{\emph{b}} \\
        \midrule
        $U_0$ & meV & 4.7 & 4.3 & 3.5 & 4.38±0.08\\
        $U$ & meV & 4.4 & 4.3 & 3.5 & 4.21±0.17 \\
        $H$ & meV & 4.4 & 4.3 & 3.5 & 4.33±0.12\\
        $G$ & meV & 5.7 & 6.0 & 5.2 & 5.87±0.06\\
        \bottomrule
        \end{tabular}
    }
    \caption{Test results on the thermodynamic targets in QM9 with XPaiNN@xTB, as compared to those of the other top models}
    \begin{tablenotes}
        \item[1] \textsuperscript{\emph{a}} Results are taken from models' original publications;
        \item[2] \textsuperscript{\emph{b}} Averaged over four models.
    \end{tablenotes}
    \label{tab:qm9-delta}
\end{table}

\newpage 
\textbf{MD17 and rMD17}. The MD17 and rMD17 datasets comprise molecular dynamics trajectories of small organic molecules. These datasets are used to assess the capability of ML models to construct potential energy surfaces (PES) with accuracy comparable to DFT calculations in their prediction of both molecular energies and atomic forces.\cite{md17, md17_2, rmd17}

Table \ref{tab:md17} and \ref{tab:rmd17} summarize test errors in terms of MAE for various models on MD17 and rMD17, respectively. As already shown by the results on QM9, XPaiNN outperforms PaiNN in almost all tasks, while the $\Delta$-ML models such as XPaiNN@xTB further improves over the XPaiNN. These findings may lead us to draw general conclusions  regarding the suitability of equivariant GNN over invariant ones for modeling high dimensional PES. Specifically, equivariant GNNs appear to offer much improved accuracy in force prediction. Furthermore, the residual PES generated by the $\Delta$-ML approach seems to exhibit more smothness, making it more amenable to accurate fitting.

\begin{table}[htbp]
    \centering
    \resizebox{\textwidth}{!}{
        \begin{tabular}{llccccccc}
        \toprule
        Molecule & & SchNet\textsuperscript{\emph{b}} & DimeNet\textsuperscript{\emph{b}} & PaiNN\textsuperscript{\emph{b}} & NequIP\textsuperscript{\emph{c}} & Equiformer\textsuperscript{\emph{c}} & XPaiNN\textsuperscript{\emph{d}} & XPaiNN@xTB\textsuperscript{\emph{d}} \\
        \midrule
        \multirow{2}{*}{Aspirin} & energy & 16.0 & 8.8 & 6.9 & 5.7 & \textbf{5.3} & 6.6±0.2 & 5.5±0.0 \\
        ~ & force & 58.5 & 21.6 & 14.7 & 8.0 & \textbf{6.6} & 12.8±0.1 & 6.6±0.1 \\
        \midrule
        \multirow{2}{*}{Ethanol} & energy & 3.5 & 2.8 & 2.7 & \textbf{2.2} & \textbf{2.2} & 2.4±0.1 & 2.2±0.0 \\
        ~ & force & 16.9 & 10.0 & 9.7 & 3.1 & \textbf{2.9} & 6.4±0.6 & 3.1±0.1 \\
        \midrule
        \multirow{2}{*}{Benzene} & energy & 3.5 & 3.4 & - & - & \textbf{2.5} & 3.7±0.2 & 3.0±0.0 \\
        ~ & force & 13.5 & 8.1 & - & - & 8.1 & \textbf{6.9±0.1} & 6.5±0.0 \\
        \midrule
        \multirow{2}{*}{Malondialdehyde} & energy & 5.6 & 4.5 & 3.9 & 3.3 & \textbf{3.2} & 3.6±0.0 & 3.2±0.0 \\
        ~ & force & 28.6 & 16.6 & 13.8 & 5.6 & \textbf{5.4} & 9.6±0.3 & 4.9±0.1 \\
        \midrule
        \multirow{2}{*}{Naphthalene} & energy & 6.9 & 5.3 & 5.0 & 4.9 & \textbf{4.4} & 5.0±0.0 & 4.9±0.0 \\
        ~ & force & 25.2 & 9.3 & 3.3 & \textbf{1.7} & 2.0 & 3.2±0.1 & 1.7±0.0 \\
        \midrule
        \multirow{2}{*}{Salicylic Acid} & energy & 8.7 & 5.8 & 4.9 & 4.6 & \textbf{4.3} & 4.8±0.0 & 4.6±0.0 \\
        ~ & force & 36.9 & 16.2 & 8.5 & \textbf{3.9} & \textbf{3.9} & 7.2±0.1 & 3.4±0.0 \\
        \midrule
        \multirow{2}{*}{Toluene} & energy & 5.2 & 4.4 & 4.1 & 4.0 & \textbf{3.7} & 4.1±0.0 & 4.0±0.0 \\
        ~ & force & 24.7 & 9.4 & 4.1 & \textbf{2.0} & 2.1 & 3.5±0.1 & 1.9±0.0 \\
        \midrule
        \multirow{2}{*}{Uracil} & energy & 6.1 & 5.0 & 4.5 & 4.5 & \textbf{4.3} & 4.5±0.0 & 4.5±0.0 \\
        ~ & force & 24.3 & 13.1 & 6.0 & \textbf{3.3} & 3.4 & 5.6±0.1 & 2.7±0.0 \\
        \bottomrule
        \end{tabular}
    }
    \caption{Test mean absolute errors on MD17.\textsuperscript{\emph{a}}}
    \begin{tablenotes}
        \item[1] \textsuperscript{\emph{a}} Unit in meV and meV/{\AA} for energy and force, respectively, the best results are in \textbf{bold}.
        \item[2] \textsuperscript{\emph{b}} \textsuperscript{\emph{c}} Results are taken from Ref.~\citenum{painn} and Ref.~\citenum{equiformer}
        \item[3] \textsuperscript{\emph{d}} Averaged over four models.
    \end{tablenotes}
    \label{tab:md17}
\end{table}

\begin{table}[htbp]
    \centering
    \resizebox{\textwidth}{!}{
        \begin{tabular}{llccccc}
        \toprule
        Molecule & & FCHL19\textsuperscript{\emph{b}} & NequIP\textsuperscript{\emph{b}} & Allegro\textsuperscript{\emph{b}} & XPaiNN\textsuperscript{\emph{c}} & XPaiNN@xTB\textsuperscript{\emph{c}} \\
        \midrule
        \multirow{2}{*}{Aspirin} & energy & 6.2 & \textbf{2.3} & \textbf{2.3} & 4.1±0.3 & 2.1±0.0 \\
        ~ & force & 20.9 & 8.2 & \textbf{7.3}& 13.5±0.1 & 6.9±0.0 \\
        \midrule
        \multirow{2}{*}{Azobenzene} & energy & 2.8 & \textbf{0.7} & 1.2 & 1.7±0.2 & 0.7±0.0 \\
        ~ & force & 10.8 & 2.9 & 2.6 & 6.0±0.1 & 3.0±0.0 \\
        \midrule
        \multirow{2}{*}{Benzene} & energy & 0.3 & \textbf{0.04} & 0.3 & 0.2±0.1 & 0.04±0.01 \\
        ~ & force & 2.6 & 0.3 & \textbf{0.2} & 0.9±0.0 & 0.3±0.0 \\
        \midrule
        \multirow{2}{*}{Ethanol} & energy & 0.9 & \textbf{0.4} & \textbf{0.4} & 1.0±0.0 & 0.4±0.0 \\
        ~ & force & 6.2 & 2.8 & \textbf{2.1} & 6.4±0.1 & 2.9±0.1 \\
        \midrule
        \multirow{2}{*}{Malondialdehyde} & energy & 1.5 & 0.8 & \textbf{0.6} & 1.6±0.2 & 0.7±0.0 \\
        ~ & force & 10.2 & 5.1 & \textbf{3.6} & 9.5±0.4 & 4.7±0.1 \\
        \midrule
        \multirow{2}{*}{Naphthalene} & energy & 1.2 & \textbf{0.2} & 0.5 & 0.7±0.1 & 0.2±0.0 \\
        ~ & force & 6.5 & 1.3 & \textbf{0.9} & 2.4±1.6 & 1.3±0.0 \\
        \midrule
        \multirow{2}{*}{Paracetamol} & energy & 2.9 & \textbf{1.4} & 1.5 & 2.4±0.1 & 1.2±0.0 \\
        ~ & force & 12.2 & 5.9 & \textbf{4.9} & 9.6±0.2 & 4.9±0.0 \\
        \midrule
        \multirow{2}{*}{Salicylic Acid} & energy & 1.8 & \textbf{0.7} & 0.9 & 1.4±0.0 & 0.6±0.0 \\
        ~ & force & 9.5 & 4.0 & \textbf{2.9} & 7.5±0.1 & 3.5±0.1 \\
        \midrule
        \multirow{2}{*}{Toluene} & energy & 1.6 & \textbf{0.3} & 0.4 & 0.7±0.1 & 0.2±0.0 \\
        ~ & force & 8.8 & \textbf{1.6} & 1.8 & 3.5±0.1 & 1.4±0.0 \\
        \midrule
        \multirow{2}{*}{Uracil} & energy & 0.6 & \textbf{0.4} & 0.6 & 0.8±0.0 & 0.4±0.0 \\
        ~ & force & 4.2 & 3.1 & \textbf{1.8} & 5.6±0.1 & 2.5±0.0 \\
        \bottomrule
        \end{tabular}
    }
    \begin{tablenotes}
        \item[1] \textsuperscript{\emph{a}} Unit in meV and meV/{\AA} for energy and force respectively, best results in \textbf{bold} except for $\Delta$-ML models.
        \item[2] \textsuperscript{\emph{b}} Results are taken from Ref.~\citenum{allegro}
        \item[3] \textsuperscript{\emph{c}} Averaged over four models.
    \end{tablenotes}
    \caption{Test mean absolute errors on rMD17. \textsuperscript{\emph{a}}}
    \label{tab:rmd17}
\end{table}

\newpage 
\textbf{Additional subsets in GMTKN55}. It is of great interest to take a further look on the performances of our models on chemical diverging systems such as charged, open-shell molecules as well as species containing untrained elements, as our training dataset does not involve these systems. For this purpose, we have selected additional subsets from GMTKN55 that contain charged, open-shell molecules as well as unseen elements to form three subsets, which complement our original benchmark. The description of the chosen subsets can be found in Table \ref{tab:gmtkn55_description}, and detailed results are provided in Table \ref{tab:gmtkn55_full}. It is worth noting that, AIQM1 model does not support this additional test due to the limitation on element types.

As can be seen, for subsets containing charged or open-shell species, direct learning model XPaiNN performs poorly due to its inability to accurately distinguish the complexities of electronic structures. On the other hand, $\Delta$-ML model XPaiNN@xTB continues to offer much improved descriptions on these systems, such as RSE43, WATER27 and IL6. This superiority shall largely to attributed to the integration of the SQM method. Performances are close between XPaiNN@xTB and OrbNet-Equi in the first two selected collections in Table \ref{tab:gmtkn55_full}. A better result of OrbNet-Equi in the third collection can be understood as its training set covers a wider range of chemical elements than our XPaiNN. For example, Si that presents in ICONF  is known to OrbNet-Equi. The same is true for B exists in various small molecules involved in reactions in NBPRC and G2RC. 

All in all, our findings here reinforce the opinions in our main text that an enhanced transferability is achieved via the $\Delta$-ML approach in decribing the chemically diverging systems. Moreover, these results hightlightthe necessity of encompassing a broad chemical space for constructing atomistic ML model of general purpose, where data-efficiency is of critical concern.

\newpage
{\small
 \renewcommand{\arraystretch}{0.75}
\begin{longtable}{ll}
    \caption{Description of benchmark subset in GMTKN55 database.}
    \label{tab:gmtkn55_description} \\
    
    \toprule
    Subset & Description \\
    \midrule
    \endfirsthead
    
    \toprule
    Subset & Description \\
    \midrule
    \endhead
    
    \bottomrule
    \multicolumn{2}{l}{\textsuperscript{\emph{a}} Subsets consist of neutral and closed-shell molecules.} \\
    \multicolumn{2}{l}{\textsuperscript{\emph{b}} Subsets contain charged but closed-shell molecules.} \\
    \multicolumn{2}{l}{\textsuperscript{\emph{c}} Subsets contain charged and open-shell molecules.} \\
    \multicolumn{2}{l}{\textsuperscript{\emph{d}} Subsets of uncharged and closed-shell molecules, but involve unknown elements.} \\
    \endfoot
    
    \multicolumn{2}{c}{Basic properties and reaction energies for small systems} \\
    AL2X6\textsuperscript{\emph{d}} & Dimerization energies of \ce{AlX3} compounds \\
    NBPRC\textsuperscript{\emph{d}} & Oligomerizations and \ce{H2} fragmentations of \ce{NH3}/\ce{BH3} systems \\
                                    & \ce{H2} activation reactions with \ce{PH3}/\ce{BH3} systems \\
    ALK8\textsuperscript{\emph{b}} & Dissociation and other reactions of alkaline compounds \\
    RC21\textsuperscript{\emph{c}} & Fragmentations and rearrangements in radical cations \\
    G2RC\textsuperscript{\emph{d}} & Reaction energies of selected G2/97 systems \\
    BH76RC\textsuperscript{\emph{c}} & Reaction energies of the BH76 set \\
    FH51\textsuperscript{\emph{a}} & Reaction energies in various (in-)organic systems \\
    TAUT15\textsuperscript{\emph{a}} & Relative energies in tautomers \\
    \midrule
    \multicolumn{2}{c}{Reaction energies for large systems and isomerization reactions} \\
    DARC\textsuperscript{\emph{a}} & Reaction energies of Diels-Alder reactions \\
    RSE43\textsuperscript{\emph{c}} & Radical-stabilization energies \\
    BSR36\textsuperscript{\emph{a}} & Bond-separation reactions of saturated hydrocarbons \\
    CDIE20\textsuperscript{\emph{a}} & Double-bond isomerization energies in cyclic systems \\
    ISO34\textsuperscript{\emph{a}} & Isomerization energies of small- and medium-sized organic molecules \\
    C60ISO\textsuperscript{\emph{a}} & Relative energies between \ce{C60} isomers\\
    PArel\textsuperscript{\emph{b}} & Relative energies in protonated isomers \\
    \midrule
    \multicolumn{2}{c}{Reaction barrier heights} \\
    BH76\textsuperscript{\emph{c}} & Barrier heights of hydrogen transfer, heavy atom transfer, \\
                                   & nucleophilic substitution, unimolecular, and association reactions \\
    BHPERI\textsuperscript{\emph{d}} & Barrier heights of pericyclic reactions \\
    BHDIV10\textsuperscript{\emph{d}} & Diverse reaction barrier heights \\
    BHROT27\textsuperscript{\emph{a}} & Barrier heights for rotation around single bonds \\
    PX13\textsuperscript{\emph{a}} & Proton-exchange barriers in \ce{H2O}, \ce{NH3}, and \ce{HF} clusters \\
    WCPT18\textsuperscript{\emph{a}} & Proton-transfer barriers in uncatalyzed and water-catalyzed reactions \\
    \midrule
    \multicolumn{2}{c}{Intermolecular noncovalent interactions} \\
    ADIM6\textsuperscript{\emph{a}} & Interaction energies of n-alkane dimers \\
    S22\textsuperscript{\emph{a}} & Binding energies of noncovalently bound dimers \\
    S66\textsuperscript{\emph{a}} & Binding energies of noncovalently bound dimers \\
    HEAVY28\textsuperscript{\emph{d}} & Noncovalent interaction energies between heavy element hydrides \\
    WATER27\textsuperscript{\emph{b}} & Binding energies in \ce{(H2O)n}, \ce{H+(H2O)n}, and \ce{OH-(H2O)n} \\
    CARBHB12\textsuperscript{\emph{d}} & Hydrogen-bonded complexes between carbene analogues and \ce{H2O}, \\
                                       & \ce{NH3}, or \ce{HCl} \\
    PNICO23\textsuperscript{\emph{d}} & Interaction energies in pnicogen-containing dimers \\
    HAL59\textsuperscript{\emph{a}} & Binding energies in halogenated dimers (incl. halogen bonds) \\
    AHB21\textsuperscript{\emph{b}} & Interaction energies in anion-neutral dimers \\
    CHB6\textsuperscript{\emph{b}} & Interaction energies in cation-neutral dimers \\
    IL16\textsuperscript{\emph{b}} & Interaction energies in anion-cation dimers \\
    \midrule
    \multicolumn{2}{c}{Intramolecular noncovalent interactions} \\
    IDISP\textsuperscript{\emph{a}} & Intramolecular dispersion interactions \\
    ICONF\textsuperscript{\emph{d}} & Relative energies in conformers of inorganic systems \\
    ACONF\textsuperscript{\emph{a}} & Relative energies of alkane conformers \\
    Amino20x4\textsuperscript{\emph{a}} & Relative energies in amino acid conformers \\
    PCONF21\textsuperscript{\emph{a}} & Relative energies in tri- and tetrapeptide conformers \\
    MCONF\textsuperscript{\emph{a}} & Relative energies in melatonin conformers \\
    SCONF\textsuperscript{\emph{a}} & Relative energies of sugar conformers \\
    UPU23\textsuperscript{\emph{b}} & Relative energies between RNA-backbone conformers \\
    BUT14DIOL\textsuperscript{\emph{a}} & Relative energies in butane-1,4-diol conformers \\
\end{longtable}
}

\begin{table}[htbp]
    \centering
    \resizebox{\textwidth}{!}{
        \begin{tabular}{lcccccc}
            \toprule
            Subset & $\omega$B97M-D3(BJ) & GFN2-xTB & GFN-xTB & OrbNet-Equi & XPaiNN & XPaiNN@xTB \\
            \midrule
            \multicolumn{7}{c}{Basic properties and reaction energies for small systems} \\
            AL2X6\textsuperscript{\emph{d}} & 3.35 & 23.17 & 24.04 & 20.13 & 138.61 & 96.94 \\
            NBPRC\textsuperscript{\emph{d}} & 2.88 & 21.55 & 22.48 & 23.78 & 64.77 & 30.73 \\
            ALK8\textsuperscript{\emph{b}} & 2.57 & 21.71 & 47.71 & 68.22 & 72.57 & 30.06 \\
            RC21\textsuperscript{\emph{c}} & 2.29 & 37.68 & 35.10 & 36.22 & 56.70 & 33.80 \\
            G2RC\textsuperscript{\emph{d}} & 2.32 & 24.31 & 32.46 & 16.20 & 27.75 & 31.96 \\
            BH76RC\textsuperscript{\emph{c}} & 2.11 & 49.19 & 56.17 & 45.99 & 50.26 & 54.51 \\
            \multicolumn{7}{c}{Reaction energies for large systems and isomerization reactions} \\
            RSE43\textsuperscript{\emph{c}} & 5.64 & 56.92 & 50.68 & 44.20 & 49.31 & 26.83 \\
            PArel\textsuperscript{\emph{b}} & 7.78 & 71.97 & 55.78 & 43.37 & 91.87 & 75.47 \\
            \multicolumn{7}{c}{Reaction barrier heights} \\
            BH76\textsuperscript{\emph{c}} & 3.85 & 59.86 & 64.16 & 56.99 & 45.79 & 53.94 \\
            BHPERI\textsuperscript{\emph{d}} & 3.31 & 27.87 & 25.38 & 10.54 & 72.92 & 16.05 \\
            BHDIV10\textsuperscript{\emph{d}} & 1.49 & 10.18 & 10.54 & 8.15 & 17.42 & 10.39 \\
            HEAVY28\textsuperscript{\emph{d}} & 12.38 & 27.83 & 29.96 & 54.33 & 60.46 & 42.59 \\
            \multicolumn{7}{c}{Intermolecular noncovalent interactions} \\
            WATER27\textsuperscript{\emph{b}} & 0.35 & 2.14 & 5.18 & 12.03 & 20.15 & 1.86 \\
            CARBHB12\textsuperscript{\emph{d}} & 2.05 & 16.89 & 6.32 & 19.69 & 16.89 & 20.92 \\
            PNICO23\textsuperscript{\emph{d}} & 3.42 & 14.70 & 31.02 & 39.03 & 95.09 & 72.55 \\
            AHB21\textsuperscript{\emph{b}} & 1.12 & 7.51 & 11.83 & 18.59 & 46.36 & 15.20 \\
            CHB6\textsuperscript{\emph{b}} & 1.94 & 11.47 & 8.37 & 23.97 & 90.08 & 13.31 \\
            IL16\textsuperscript{\emph{b}} & 0.58 & 2.25 & 2.97 & 2.55 & 31.79 & 3.08 \\
            \multicolumn{7}{c}{Intramolecular noncovalent interactions} \\
            ICONF\textsuperscript{\emph{d}} & 3.06 & 28.35 & 45.73 & 29.76 & 65.08 & 51.39 \\
            UPU23\textsuperscript{\emph{b}} & 6.25 & 28.87 & 12.31 & 10.99 & 17.81 & 18.10 \\
            \midrule
            WTMAD-2 \emph{b} & 3.09 & 21.47 & 18.73 & 21.20 & 44.58 & 22.02 \\
            WTMAD-2 \emph{c} & 3.81 & 54.49 & 55.75 & 49.25 & 48.82 & 44.69 \\
            WTMAD-2 \emph{d} & 4.50 & 22.85 & 27.65 & 27.22 & 60.14 & 39.35 \\
            \bottomrule
        \end{tabular}
    }
    \begin{tablenotes}
        \item[1] \textsuperscript{\emph{b}} Subsets contain charged but closed-shell molecules.
        \item[2] \textsuperscript{\emph{c}} Subsets contain charged and open-shell molecules.
        \item[3] \textsuperscript{\emph{d}} Subsets contain uncharged and closed-shell molecules, but involve unknown elements.
    \end{tablenotes}
    \caption{WTMAD-2 for additional subsets in GMTKN55. Unit in kcal/mol.}
    \label{tab:gmtkn55_full}
\end{table}

\newpage 
\textbf{ROT34}. Table \ref{tab:rot34} summarizes the RMSD for each molecule between reference structures and the optimized structures sourced by different methods. Optimized structures of GFN-xTB, GFN2-xTB, ANI-2x and OrbNet-Equi are obtained from Ref.~\citenum{orbnet-equi}, while the others can be found in the attached zip file. RMSDs are calculated using the quaternions method\cite{rmsd}.
\begin{table}[htbp]
    \centering
    \resizebox{\textwidth}{!}{
        \begin{tabular}{lcccccccc}
            \toprule
            Compound & $\omega$B97M-D3(BJ) & GFN-xTB & GFN2-xTB & ANI-2x & OrbNet-Equi & XPaiNN & XPaiNN@xTB \\
            \midrule
            Ethynyl-cyclohexane & 0.006 & 0.012 & 0.016 & 0.015 & 0.011 & 0.008 & 0.011 \\
            Isoamyl-acetate & 0.035 & 0.692 & 0.500 & 0.069 & 0.141 & 0.058 & 0.053 \\
            Diisopropylketone & 0.015 & 0.057 & 0.043 & 0.062 & 0.020 & 0.052 & 0.059 \\
            Bicyclo[2.2.2]octadiene & 0.005 & 0.031 & 0.029 & 0.035 & 0.013 & 0.026 & 0.016 \\
            Triethylamine & 0.021 & 0.119 & 0.138 & 0.071 & 0.013 & 0.055 & 0.055 \\
            Vitamin C & 0.016 & 0.073 & 0.071 & 0.143 & 0.054 & 0.058 & 0.028 \\
            Serotonin & 0.024 & 0.031 & 0.053 & 0.077 & 0.036 & 0.018 & 0.026 \\
            Aspirin & 0.072 & 0.126 & 0.192 & 0.229 & 0.037 & 0.118 & 0.112 \\
            Cassyrane & 0.012 & 0.096 & 0.084 & 0.289 & 0.059 & 0.075 & 0.038 \\
            Limonene & 0.020 & 0.075 & 0.029 & 0.031 & 0.031 & 0.015 & 0.023 \\
            Lupinine & 0.011 & 0.054 & 0.046 & 0.033 & 0.038 & 0.026 & 0.020 \\
            Proline derivative & 0.016 & 0.097 & 0.038 & 0.070 & 0.026 & 0.033 & 0.031 \\
            \midrule
            Average & 0.021 & 0.122 & 0.103 & 0.094 & 0.040 & 0.045 & 0.039 \\
            \bottomrule
        \end{tabular}
    }
    \caption{RMSD of different methods for each molecule in ROT34. Unit in \AA}
    \label{tab:rot34}
\end{table}

\textbf{Computation Efficiency}. To compare method efficiency, we document the time consumption of methods used for calculating the reaction enthalpy of reaction \textbf{a} in Fig.4 in the main text. The results are listed in Table \ref{tab:vibration}. As can be seen, our model is significantly faster than the DFT method while maintaining a close accuracy.
\begin{table}[htbp]
    \centering
    \resizebox{\textwidth}{!}{
        \begin{tabular}{lccccc}
            \toprule
            Reaction & $\omega$B97M-D3(BJ) & GFN2-xTB & XPaiNN & XPaiNN@xTB \\
            \midrule
            Total CPU Time & $3.20\times10^5$ & 15.6 & 107 & 109 \\
            \bottomrule            
        \end{tabular}
    }
    \caption{CPU time for reaction \textbf{a}. calculated by different methods. Unit in second.}
    \label{tab:vibration}
\end{table}

Detailed results for the \textbf{S66x8}, \textbf{Torsion}, and \textbf{BH9} datasets are provided in the attached zip file.

\newpage

\bibliography{references}